\documentclass[aps,prx,10pt,twocolumn,superscriptaddress,nofootinbib,preprintnumbers,showpacs]{revtex4-2}
\usepackage{amsmath,amsfonts, amssymb, amsthm, dsfont}
\usepackage[T1]{fontenc}
\usepackage{yfonts}
\usepackage{bm}
\usepackage{bbm}
\usepackage{mathrsfs}
\usepackage{graphicx}
\usepackage{verbatim}
\usepackage[bookmarks=true,colorlinks=true,linkcolor=blue, urlcolor=blue, citecolor=blue]{hyperref}
\hypersetup{linktocpage}
\usepackage{wasysym}
\usepackage[dvipsnames]{xcolor}
\usepackage{bbold}
\usepackage{braket}
\usepackage{diagbox} 
\usepackage{environ} 
\usepackage{relsize} 
\usepackage[caption=false]{subfig} 
\usepackage{tikz}
\usepackage{ulem}

\NewEnviron{eqs}{%
\begin{equation}\begin{split}
    \BODY
\end{split}\end{equation}
}

\newcommand{\comments}[1]{}

\renewcommand{\cal}[1]{\mathcal{#1}}

\def\sD{\mathsf{D}}

\DeclareMathOperator{\Tr}{Tr}
\DeclareMathOperator{\re}{Re}
\DeclareMathOperator{\im}{Im}

\theoremstyle{plain}
\newtheorem*{theorem*}{Theorem}

\makeatletter
\def\l@subsubsection#1#2{}
\makeatother

\begin{document}

\title{Corner Charge Fluctuations and Many-Body Quantum Geometry}
\author{Xiao-Chuan Wu}
\affiliation{Kadanoff Center for Theoretical Physics \& Enrico Fermi Institute, University of Chicago, Chicago, IL 60637, USA}
\author{Kang-Le Cai}
\affiliation{Department of Physics, Yale University, New Haven, Connecticut 06511, USA}
\author{Meng Cheng} 
\affiliation{Department of Physics, Yale University, New Haven, Connecticut 06511, USA}
\author{Prashant Kumar}
\affiliation{Kadanoff Center for Theoretical Physics \& Enrico Fermi Institute, University of Chicago, Chicago, IL 60637, USA}
\affiliation{Department of Physics, Indian Institute of Technology Bombay, Mumbai 400076, India}


\begin{abstract}

In many-body systems with U(1) global symmetry, the charge fluctuations in a subregion reveal important insights into entanglement and other global properties. For subregions with sharp corners, bipartite fluctuations have been predicted to exhibit a {\it universal shape dependence} on the corner angle in certain quantum phases and transitions, characterized by a ``universal angle function'' and a ``universal coefficient.'' However, we demonstrate that this simple formula is insufficient for charge insulators, including composite fermi liquids. In these systems, the corner contribution may depend on the corner angle, subregion orientation, and other microscopic details. We provide an infinite series representation of the corner term, introducing {\it orientation-resolved universal angle functions} with their {\it non-universal coefficients}. In the small-angle limit or under orientation averaging, the remaining terms’ coefficients are fully determined by the {\it many-body quantum metric}, which, while not universal, adheres to both a universal topological lower bound and an energetic upper bound. We also clarify the conditions for bound saturation in (anisotropic) Landau levels, leveraging the generalized Kohn theorem and holomorphic properties of many-body wavefunctions. We find that a broad class of fractional quantum Hall wavefunctions, including unprojected parton states and composite-fermion Fermi sea wavefunctions, saturates the bounds.

\end{abstract}


\maketitle

\tableofcontents

\section{Introduction}

Understanding the universal properties of quantum phases and phase transitions is a primary goal of condensed matter physics. Many physical quantities are governed by universal numbers that are insensitive to the microscopic details of the systems. A well-known example is the Hall conductivity of integer or fractional quantum (anomalous) Hall insulators, which is determined by a topological invariant known as the many-body Chern number~\cite{tknn,tknn2}. Similarly, various dynamical and static response functions at conformal quantum critical points are determined by the universality class of the transition (see e.g.~\cite{sachdevbook1,bootstrapRMP} for a review).


In recent years, with the aid of the concept of quantum geometry~\cite{QuanGeom1980}, it has become apparent that some physical quantities, while typically taking non-universal values in gapped phases, are actually universally bounded by the topological properties. There is a class of examples involving the Chern number that stem from the positive semi-definiteness of the quantum geometric tensor~\cite{Roy_bound}, which has quantum metric and Berry curvature as its real and imaginary parts, respectively. A notable example is the lower bound on the superfluid weight in multi-band superconductors~\cite{SF_bound1,SF_bound2}, which has applications in twisted bilayer graphene~\cite{SF_bound_rev,SF_bound_tbg1,SF_bound_tbg2}. Additionally, it was recently shown that the energy gap in Chern insulators is subject to a fundamental upper bound~\cite{Fu2024-1,Fu2024-2,Fu2024-3}. Besides bounds related to Chern numbers, an earlier foundational result discussed in Ref. \cite{Haldane2009}, called the Haldane bound, bounds the measure of quadrupolar fluctuations in fractional QH systems. This result links the quantum metric in the parameter space of linear area-preserving diffeomorphisms to the Hall viscosity. It was recently shown that the universal bound is not generally saturated~\cite{Kumar2024}.

On a different front, recent studies have emphasized the growing importance of non-local characterizations of phases of matter, such as symmetry disorder operators and bipartite fluctuations of local observables\footnote{In systems with global $\textrm{U}(1)$ symmetry, the bipartite fluctuations of the conserved $\textrm{U}(1)$ charges can be identified as the $\textrm{U}(1)$ disorder operator under the small-angle limit~\cite{wulog,chengdisop}.}. This is partly motivated by the conceptual developments of  topological defects and generalized symmetries ~\cite{Nussinov2009SymTO,Gaiotto2015GeneSym,gene_sym_review1,gene_sym_review2,wulog,chengdisop}, as well as the experimental proposals for measuring entanglement entropy~\cite{EEBF1,EEBF2,EEBF3,EEBF4,EEBF6,EEBF7,EEBF9}. The disorder operators, or bipartite fluctuations, are associated with a spatial submanifold of the entire system and can exhibit {\it universal shape dependence} in certain scenarios. For instance, when the boundary of the subsystem is not smooth and includes a sharp corner, as depicted in FIG.~\ref{fig:_corner}, a subleading corner contribution is generally present. This occurs in systems such as non-interacting Dirac systems~\cite{Dirac_log}, conformal field theories (CFTs)~\cite{wulog,chengdisop,uni_corner}, gapped quantum Hall (QH) states~\cite{uni_corner}, composite Fermi liquid states~\cite{BF_CFS}, and certain non-Fermi liquids~\cite{BF_CFS} associated with exotic phase transitions out of Fermi liquids. Remarkably, the corner contributions in various systems share the same ``super-universal angle function''~\cite{uni_corner} (as given in Eq.~\eqref{eq:_angle_function_0}), with the coefficient determined by a ``universal number'' specific to each phase of matter. Corner contributions in several examples of critical points and QH states have been examined numerically using Monte Carlo simulations~\cite{uni_corner,Mengdisop2,Mengdisop3,Mengdisop4,Mengdisop5,CFLEE3}.

In this work, we reveal an intriguing relationship between the corner contribution of bipartite fluctuations and the many-body quantum geometry defined by adiabatic flux insertion (or twisted boundary conditions). We highlight that in charge insulators\footnote{Here, a charge insulator is defined by finite polarization fluctuations (or a finite localization tensor)~\cite{Kohn1964,SWM2000,Resta2002Rev,Resta2011Rev}.}, the corner contribution is generally {\it non-universal}. Even in the special case of Landau levels (LLs) with continuous translational and rotational symmetries, certain conditions must be met for the coefficient to match the previously predicted universal number~\cite{uni_corner,BF_CFS}, even though the ``super-universal angle function'' still holds. More generally, for anisotropic continuous systems or long-wavelength descriptions of lattice models, we demonstrate that the corner contribution consists of a series of universal corner functions along with their non-universal coefficients. These corner functions typically depend on both the corner angle and the absolute orientation of the subsystem. Furthermore, we show that, in the small-angle limit or under orientation averaging, only a few terms remain in the series summation. The full corner contribution is then governed by a universal angle dependence and a coefficient determined entirely by the {\it many-body quantum metric} (or polarization fluctuations), as described in Eq.~\eqref{eq:_BF_small-angle} and Eq.~\eqref{eq:_BF_avg}. {We confirm these expectations by numerically computing the corner term in tight-binding lattice models and through analytical results for an exactly solvable case. In the context of non-interacting fermions,  
the small-angle limit was first discussed in Ref.~\cite{corner_QM}.} 


In charge insulators, the quantum metric is known to satisfy a lower bound set by the Chern number~\cite{Roy_bound} and an upper bound related to the energy gap~\cite{SWM2000}. Recently, the topological bound has also been understood through optical absorption, where the absorbed power must be non-negative~\cite{Fu2024-3}. In this work, we refine our understanding by providing specific conditions for bound saturation. We establish a generalized version of the Kohn theorem~\cite{Kohn1964} for systems invariant under continuous translations and Galilean boosts without assuming continuous rotational symmetry. The theorem guarantees that the quantum metric is fixed by the filling factor and the (unimodular and generally anisotropic) mass tensor in the UV and is independent of the interactions. When considering the coefficient in Eq.~\eqref{eq:_BF_avg}, the energetic upper bound is always saturated, while the lower bound is saturated only when the mass tensor is isotropic. By examining polarization fluctuations in anisotropic Landau levels, we additionally provide an understanding of bound saturation from the perspective of the holomorphic properties of many-body wavefunctions in a wider set of scenarios.

The outline of the rest of the paper is as follows. In Sec.~\ref{sec:_Corner_Rev}, we begin by introducing existing results from Refs.~\cite{Dirac_log,wulog,uni_corner,chengdisop,BF_CFS} and summarizing some of our key findings, emphasizing their relation to many-body quantum geometry. In Sec.~\ref{sec:_Dis_Sym}, we present a general framework for understanding corner charge fluctuations in both continuous and lattice models. Sec.~\ref{sec:_Chern_insulators} explores examples of lattice models, providing a comparison between analytical formulas and numerical results. In Sec.~\ref{sec:_Ani_LL}, we present general results on (anisotropic) Landau levels, with a particular focus on understanding the bound saturation of the quantum metric. Composite Fermi liquids (CFLs) are also addressed in Sec.~\ref{sec:_CFL}. Finally, in Sec.~\ref{sec:_Summary}, we summarize our findings and suggest interesting future research directions.

\section{Corner Charge Fluctuations}
\label{sec:_Corner_Rev}

\subsection{Preliminaries}
\label{subsec:_Pre}


For any quantum many-body system with a global U(1) symmetry, the disorder operator is defined as
\begin{flalign}
\mathcal{U}_{\textrm{A}}(\alpha)=\exp\left(\mathtt{i}\alpha\int_{\textrm{A}}\rho\right).
\label{eq:_dis_op}
\end{flalign}
Here, $\alpha$ is a real-valued parameter, $\rho$ represents the globally conserved charge density, and $\textrm{A}$ is a spatial subregion within the total system. The expectation value $\langle\mathcal{U}_{\textrm{A}}(\alpha)\rangle$ can be viewed as a generating functional, the $m$-th cumulant of which is given by 
\begin{flalign}
\mathcal{N}_{\textrm{A}}^{[m]}=\lim_{\alpha\rightarrow0}(-\mathtt{i}\partial_{\alpha})^{m}\log\langle\mathcal{U}_{\textrm{A}}(\alpha)\rangle.
\label{eq:_cumulant}
\end{flalign}

\begin{figure}
\includegraphics[width=0.35\textwidth]{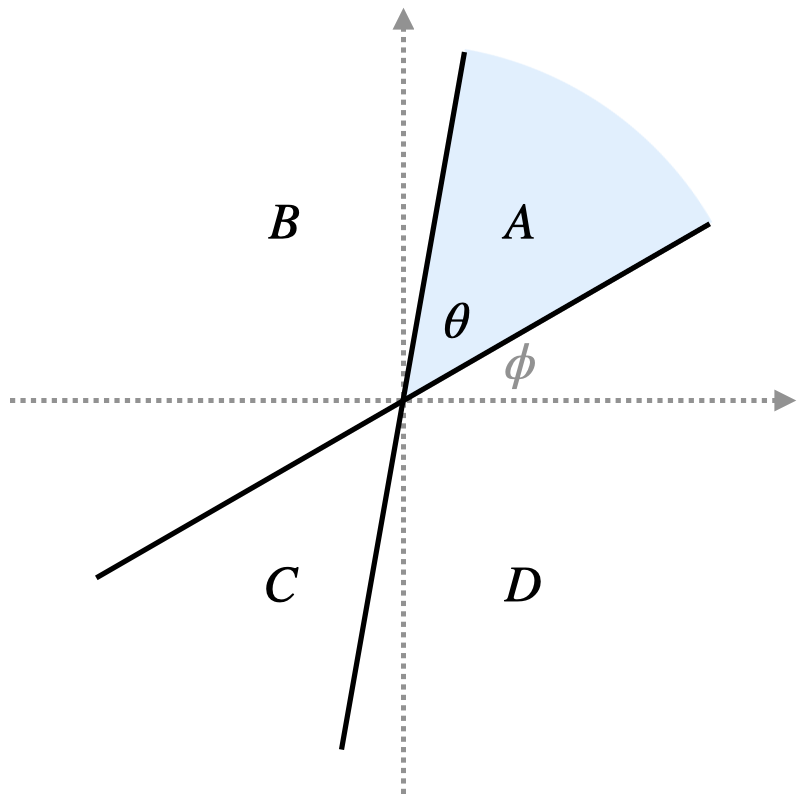}
\caption{The real-space subsystem A (shaded area) with only one corner angle $\theta$. The absolute orientation of A is parameterized by another angle $\phi$, and the linear size of A is denoted by $L_{\textrm{A}}$. The regions B, C, and D are introduced to cancel out the boundary law and isolate the corner contribution.} \label{fig:_corner}
\end{figure}

The second cumulant $\mathcal{N}_{\textrm{A}}^{[2]}$ is called the bipartite fluctuations. In recent years, there has been significant theoretical progress in understanding the behaviors of $\langle\mathcal{U}_{\textrm{A}}(\alpha)\rangle$ and $\mathcal{N}_{\textrm{A}}^{[2]}$ in two-dimensional many-body systems {\it with continuous translation and rotational symmetries}, particularly concerning subregions with a corner, such as the one depicted in FIG.~\ref{fig:_corner}. 

Specifically, for conformally invariant quantum critical points~\cite{Dirac_log,wulog,uni_corner,chengdisop}, the scaling behavior is given by 
\begin{flalign}
\mathcal{N}_{\textrm{A}}^{[2]}=\#L_{\rm A}-\frac{C_{J}}{2}f_{0}(\theta)\log(L_{\rm A})+\ldots
\label{eq:_BF_CFT}
\end{flalign}
where $L_{\rm A}$ is the linear size of $A$, $\#$ is a non-universal number depending on the UV cut-off and the angle dependence is given by
\begin{equation} 
f_{0}(\theta)=1+(\pi-\theta)\cot(\theta).
\label{eq:_angle_function_0}
\end{equation}
The overall coefficient $C_{J}$, known as the current central charge~\cite{bootstrapRMP}, is a universal number associated with the critical point and relates to the longitudinal conductivity via $\sigma^{xx} = \frac{\pi^{2}}{2} C_{J}$. In unitary CFTs, one generally has $C_{J}>0$. 

More recently, Ref.~\cite{BF_CFS} pointed out that Eq.~\eqref{eq:_BF_CFT} holds for a class of non-conformal continuous transitions~\cite{senthilmit1,senthilmit2,frac_mit,cflfl2,cflfl3} out of Landau Fermi liquids, where the universal coefficient $C_{J}$ is associated with the predicted universal resistivity jump at the critical point.

Additionally, Ref.~\cite{uni_corner} showed that for an isotropic system 
\begin{flalign}
\mathcal{N}_{\textrm{A}}^{[2]}=\#L_{\rm A}-\frac{C_{0}}{\pi} f_{0}(\theta)+\ldots
\label{eq:isotropic_corner}
\end{flalign}
where $\#$ is again non-universal, $C_{0}$ is determined by the static structure factor: $S(\boldsymbol{q})\approx C_{0} |\boldsymbol{q}|^2$. We see that the angle dependence is governed by the ``super-universal'' formula Eq.~\eqref{eq:_angle_function_0}.
Based on the wavefunctions for isotropic integer QH and Laughlin states, Ref.~\cite{uni_corner} showed that in these states $C_{0}$ appears to be universal:
\begin{equation}
  C_{0}=\frac{|\sigma^{xy}|}{2},  
  \label{eq:_BF_QH}
\end{equation} 
where $\sigma^{xy}$ is the DC Hall conductivity. Ref.~\cite{BF_CFS} recently suggested that the same result also applies to certain CFLs based on an isotropic field theory~\cite{HLR1993}\footnote{The leading-order boundary law in Eq.~\eqref{eq:_BF_QH} for CFLs was independently confirmed in Ref.~\cite{disop_FS}. Both the boundary law and the corner term were later verified through Monte Carlo simulations based on wavefunctions in Ref.~\cite{CFLEE3}.}.



In this paper, we take a closer look at the corner charge fluctuations. In particular, we examine how Eq.~\eqref{eq:isotropic_corner} as well as Eq.~\eqref{eq:_BF_QH} are modified in  {more general situations}. 
  
 These issues will be discussed in detail in the following sections. Here we start with a few general facts about corner contributions. In general, we expect that
\begin{equation}
    \cal{N}_{\rm A}^{[2]}=\# L_{\rm A} - \gamma + \cdots.
\end{equation}
Here $\gamma$ is a constant that generically depends on the sharp corners of the region $\rm A$. We parametrize a sharp corner by two angles: the opening angle $\theta$ of the corner, and the absolute orientation $\phi$ (see Fig.~\ref{fig:_corner} for illustration). Then we express $\gamma$ as
\begin{equation}
    \gamma=\sum_i \gamma(\phi_i, \theta_i),
\end{equation}
where the sum is over the local contributions from all sharp corners. If the system enjoys continuous rotation symmetry, then $\gamma$ is independent of $\phi$: $\gamma=\sum_i \gamma(\theta_i)$. In this case, Ref.~\cite{uni_corner} showed that $\gamma(\theta)$ is proportional to the angle function Eq.~\eqref{eq:_angle_function_0}. However, it is important to note that the coefficient is not always given by $\sigma^{xy}$, as it is in Eq.~\eqref{eq:_BF_QH}. A simple counterexample will be mentioned in Sec.~\ref{subsec:_FQHE}. 

In the following sections, we systematically study $\gamma(\phi, \theta)$, analyzing both its universal and non-universal features. In Sec.~\ref{subsec:_Fourier}, we will demonstrate that when $\gamma(\phi, \theta)$ is expanded in harmonics of $\phi$, each component is a higher-order universal function of $\theta$, generalizing the super-universal function Eq.~\eqref{eq:_angle_function_0}. For more details, see Eq.~\eqref{eq:_BF_corner_expd} and Eq.~\eqref{eq:_angle_functions}. The intriguing properties of the coefficients will also be discussed.



\subsection{Quantum Geometry}
\label{subsec:_bound}





In this subsection, we summarize some of our key results from later sections, particularly emphasizing their connections to many-body quantum geometry. 

By definition, the bipartite fluctuations are determined by the equal-time density-density correlation $S(\boldsymbol{r})$
\begin{flalign}
\mathcal{N}_{\textrm{A}}^{[2]}=\int_{\textrm{A}}\textrm{d}^{2}\boldsymbol{r}_{1}\int_{\textrm{A}}\textrm{d}^{2}\boldsymbol{r}_{2}\,S_{+}(\boldsymbol{r}_{1}-\boldsymbol{r}_{2}),
\label{eq:_BF_1}
\end{flalign}
where $S_{+}(\boldsymbol{r})=(S(\boldsymbol{r})+S(-\boldsymbol{r}))/2$ denotes the inversion-symmetric component of $S(\boldsymbol{r})$. The static structure factor $S(\boldsymbol{q})=(1/V)\langle\rho_{\boldsymbol{q}}\rho_{-\boldsymbol{q}}\rangle$ is given by the Fourier transformation of $S(\boldsymbol{r})$, where $V=\int\textrm{d}^{2}\boldsymbol{r}$ denotes the total volume of the system. It is clear that the large-scale real-space scaling of $\mathcal{N}_{\textrm{A}}^{[2]}$ is determined by the long-wavelength behavior of $S_{+}(\boldsymbol{q})=(S(\boldsymbol{q})+S(-\boldsymbol{q}))/2$. Thus, we are motivated to consider the expansion\footnote{This analytical expansion applies to gapped systems. However, gapless models of charge insulators, such as CFLs, may exhibit non-analytical higher-order terms like  $|\boldsymbol{q}|^{3}\log(1/|\boldsymbol{q}|)$. See Sec.~\ref{sec:_CFL} for more details.}
\begin{flalign}
S_{+}(\boldsymbol{q}) = \mathsf{S}_{2}^{ab} q_{a} q_{b} + \mathscr{O}(|\boldsymbol{q}|^{4}),
\label{eq:_SSF_1}
\end{flalign}
where $\mathsf{S}_{2}^{ab}$ is the symmetric part of the localization tensor (i.e., polarization fluctuations)~\cite{Kohn1964,SWM2000,Resta2002Rev,Resta2011Rev}. Recently, $\mathsf{S}_{2}^{ab}$ has also been referred to as the ``quantum weight'' in Refs. \cite{Fu2024-1,Fu2024-2,Fu2024-3}. Importantly, a general criterion that distinguishes charge insulators from metals is given by $\mathsf{S}_{2}^{ab}$, which is finite in insulators and divergent in conventional metals. This can be understood through the Souza-Wilkens-Martin (SWM) sum rule~\cite{SWM2000} 
\begin{flalign}
\mathsf{S}_{2}^{ab}=\int_{0}^{+\infty}\frac{\textrm{d}\omega}{\pi}\frac{\re\sigma_{+}^{ab}(\omega)}{\omega},
\label{eq:_SWM_sum}
\end{flalign}
where we use $\sigma_{\pm}^{ab}=(\sigma^{ab}\pm\sigma^{ba})/2$ to denote the longitudinal and Hall conductivities respectively\footnote{In Eq.~\eqref{eq:_SWM_sum}, the conductivity $\sigma^{ab}$ is defined by the response to the external electric field. To obtain the conductivity corresponding to the current induced by the total (or screened) electric field, $\sigma^{ab}$ needs to be rescaled by the dielectric constant.}. The sum rule is a direct consequence of the Ward identity relating density response and current response. Clean metals\footnote{Here, we include Landau fermi liquids as well as non-fermi liquids within the Hertz-Millis paradigm~\cite{Lu1_2}.} have a nonzero Drude weight $D^{ab}$ such that\footnote{The notation $A\supset B$ means that $B$ is a term included in the full expression of $A$.}
\begin{flalign}
\sigma_{+}^{ab}(\omega)\supset D^{ab}\left(\delta(\omega)+\frac{\mathtt{i}}{\pi\omega}\right),
\end{flalign}
which predicts a $1/\omega$ divergence in $\mathsf{S}_{2}^{ab}$ based on Eq.~\eqref{eq:_SWM_sum}. In fermi liquids, the identical scaling of frequency $\omega$ and momentum $q$ implies $S(q)\sim q$, which is consistent with the geometric interpretation~\cite{BF_CFS} based on the LU(1) anomaly~\cite{LU1_1}. On the other hand, when the Drude weight vanishes, $\mathsf{S}_{2}^{ab}$ has the chance of being finite. As discussed below, this finite value is associated with a finite corner contribution to bipartite fluctuations. It is important to note that in a clean system, a vanishing Drude weight does not necessarily indicate an incompressible state. As we will discuss in Sec.~\ref{sec:_CFL}, the half-filled Landau level~\cite{Kohn1964,HLR1993,cfl_drude} serves as an example of this\footnote{Other examples of compressible states with vanishing Drude weight include the quantum Lifshitz model~\cite{critical_drag} and the ``bad metal'' constructed from a vortex fermi-liquid state~\cite{xu_bad}.}. 

A simple dimensional analysis of the conductivity reveals that $\sigma^{ab}\sim(e^{2}/2\pi\hbar)L^{2-d}$, where $L$ represents a real-space length scale, and $d$ is the spatial dimension. Therefore, according to Eq.~\eqref{eq:_SWM_sum}, both $\sigma^{ab}$ and $\mathsf{S}_{2}^{ab}$ can be dimensionless universal numbers (in the unit of $e^{2}/2\pi\hbar$) only when $d=2$. For convenience, we set $e=1$ and $\hbar=1$ in most parts of the paper.

As reviewed in App.~\ref{app:_Quan_Geom}, there is a standard relationship~\cite{SWM2000,Resta2002Rev,Resta2011Rev,Fu2024-1} between the localization tensor $\mathsf{S}_{2}^{ab}$ and the many-body quantum geometry defined by adiabatic flux insertion (or twisted boundary conditions). For two-dimensional systems with $D^{ab}=0$, $\mathsf{S}_{2}^{ab}$ is directly identified as the following quantity
\begin{flalign}
g^{ab}=\frac{1}{N_{\textrm{g}}}\sum_{n=1}^{N_{\textrm{g}}}\mathcal{G}_{nn}^{ab},\label{eq:g_deg_gs}
\end{flalign}
where $\mathcal{G}_{nm}^{ab}$ represents the (non-abelian) quantum metric associated with the possibly degenerate ground states labeled by $n$ and $m$, with $1\leq n,m\leq N_{\textrm{g}}$. If the ground state is unique (i.e., $N_{\textrm{g}}=1$), $g^{ab}$ reduces to the abelian quantum metric. The SWM sum rule Eq.~\eqref{eq:_SWM_sum} establishes a fluctuation-dissipation theorem that relates excited states to the properties of ground-state wavefunctions on the manifold of twisted boundary conditions. 

When $\mathsf{S}_{2}^{ab}$ is finite, we will use the terminologies polarization fluctuations, localization tensor, quantum weight, and quantum metric interchangeably.

A key result for (generally anisotropic) continuous models as well as lattice models\footnote{For lattice models, we assume that the direction with $\phi=0$ is always aligned with one of the primitive vectors of the lattice.} is about the small-$\theta$ singularity of bipartite fluctuations concerning the geometry depicted in FIG.~\ref{fig:_corner}. We find the {\it orientation-resolved} corner term generally satisfies 
\begin{flalign}
\gamma(\phi,\theta\rightarrow0)\approx\frac{\mathsf{S}_{2}^{ab}}{\theta}\hat{\tau}_{a}(\phi)\hat{\tau}_{b}(\phi),
\label{eq:_BF_small-angle}
\end{flalign}
where $\hat{\tau}(\phi)=(-\sin\phi,\cos\phi)$ is the unit vector perpendicular to the lower edge of the subregion. When the system exhibits $C_{3}$, $C_{4}$, or $C_{6}$ discrete rotational symmetry, the quantum metric becomes isotropic, meaning $\mathsf{S}_{2}^{ab}=\mathsf{S}_{2}\delta^{ab}$. As a result, Eq.~\eqref{eq:_BF_small-angle} simplifies to $\mathsf{S}_{2}/\theta$, where $\mathsf{S}_{2}=\Tr(\mathsf{S}_{2}^{ab})/2=\sqrt{\det(\mathsf{S}_{2}^{ab})}$. 


For continuous systems, we also define the {\it orientation-averaged} corner term as follows
\begin{flalign}
\bar{\gamma}(\theta)=\int_{0}^{2\pi}\frac{\textrm{d}\phi}{2\pi}\gamma(\phi,\theta).
\label{eq:_BF_avg_def}
\end{flalign}
As demonstrated in Sec.~\ref{sec:_Dis_Sym} and Sec.~\ref{sec:_Ani_LL}, the final result resembles Eq.~\eqref{eq:_BF_QH}, but with the coefficient fully determined by the many-body quantum metric  
\begin{flalign}
\bar{\gamma}(\theta)=\frac{\Tr (\mathsf{S}_{2}^{ab})}{2\pi}(1+(\pi-\theta)\cot\theta),
\label{eq:_BF_avg}
\end{flalign}
where we define the trace with respect to the flat metric, i.e. $\Tr (\mathsf{S}_{2}^{ab}) = \delta_{ab}\mathsf{S}_{2}^{ab}$. We once again find the small-$\theta$ limit is governed by $\mathsf{S}_{2}^{ab}$
\begin{flalign}
\bar{\gamma}(\theta\rightarrow 0)\approx\frac{\Tr (\mathsf{S}_{2}^{ab})}{2}\frac{1}{\theta}.
\label{eq:_BF_avg_small-angle}
\end{flalign}
This result is also applicable to continuous systems without discrete rotational symmetry.

According to Refs.~\cite{Roy_bound,SWM2000,Resta2002Rev,Resta2011Rev,Fu2024-1} (also see App.~\ref{app:_Quan_Geom} for a brief overview), the quantum metric $\mathsf{S}_2^{ab}$ is subject to a topological lower bound due to the positive semi-definiteness of the quantum geometric tensor, which is succinctly expressed as $\mathsf{S}_2^{ab} - \frac{\mathtt{i}}{2} \sigma^{xy} \epsilon^{ab} \geq 0$. Additionally, the standard $f$-sum rule leads to an energetic upper bound~\cite{SWM2000,Resta2002Rev,Resta2011Rev,Fu2024-1}, which, in general, should be represented by the positive semi-definiteness of the matrix $\frac{1}{2\Delta}\mathsf{D}_{2}^{ab}-\mathsf{S}_{2}^{ab}\geq0$, where
\begin{flalign}
\mathsf{D}_{2}^{ab}=\lim_{A\rightarrow0}\frac{1}{V}\left\langle \!\frac{\partial^{2}H}{\partial A_{a}\partial A_{b}}\!\right\rangle
\end{flalign}
comes from the diamagnetic term of the current response, $\Delta$ denotes the energy gap of the many-body Hamiltonian $H$, $V$ is the volume of the system, $\boldsymbol{A}$ represents the background U(1) field\footnote{In quantum Hall systems, $\boldsymbol{A}$ should be understood as the probe field in addition to the gauge field representing the background magnetic field.}, and $\sigma^{xy}$ is the DC Hall conductivity. Therefore, the coefficient in Eq.~\eqref{eq:_BF_avg} satisfies 
\begin{flalign}
\frac{|\sigma^{xy}|}{2}\leq\frac{\Tr(\mathsf{S}_{2}^{ab})}{2}\leq\frac{\Tr(\mathsf{D}_{2}^{ab})}{4\Delta}.
\label{eq:_TrS2_bounds}
\end{flalign}

The topological lower bound was also obtained in Ref. \cite{Fu2024-3}. Here we elucidate its connection with quantum metric. As will be shown below, the bound is saturated in quantum Hall systems under certain conditions, as well as certain non-interacting band insulators.

As we will explain in Sec.~\ref{subsec:_Kohn}, the upper bound is saturated for systems invariant under Galilean boosts and continuous translations. If continuous rotational symmetry is also imposed, making the system invariant under the full Galilean group, then the lower and upper bounds become identical and must be saturated. We offer two perspectives to understand this result: one based on linear response theory using a generalized Kohn theorem, which we prove in App.~\ref{app:_Kohn_th}, and the other based on wavefunction holomorphicity, as discussed in Sec.~\ref{subsec:_FQHE}. In more general situations, such as systems on a lattice or with non-parabolic kinetic energy, the bounds are typically not expected to be saturated. Simple examples of this will be mentioned in Sec.~\ref{subsec:_FQHE} and Sec.~\ref{sec:_Chern_insulators}.

In addition to the application to corner charge fluctuations in Eq.~\eqref{eq:_BF_small-angle}, Eq.~\eqref{eq:_BF_avg}, and Eq.~\eqref{eq:_BF_avg_small-angle}, where the universal bounds are given in Eq.~\eqref{eq:_TrS2_bounds}, there are other related forms. In terms of determinants, the bounds can be written as
\begin{align}
    \frac{|\sigma^{xy}|}{2}\leq\sqrt{\det{(\mathsf{S}_{2}^{ab})}}\leq\frac{1}{2\Delta}\sqrt{\det{(\mathsf{D}_{2}^{ab})}}.
    \label{eq:_detS2_bounds}
\end{align}
Sometimes, the system is associated with an intrinsic unimodular metric $\eta_{ab}$, which can generally be anisotropic. The modified trace $\overline{\textrm{Tr}}(\mathsf{S}_{2}^{ab})\equiv\eta_{ab}\mathsf{S}_{2}^{ab}$ satisfies
\begin{align}
    \frac{|\sigma^{xy}|}{2}\leq\frac{\overline{\Tr}(\mathsf{S}_{2}^{ab})}{2}\leq\frac{\overline{\Tr}(\mathsf{D}_{2}^{ab})}{4\Delta},
    \label{eq:_mTrS2_bounds}
\end{align}
which turns out to be useful in Sec.~\ref{sec:_Ani_LL} for anisotropic Landau levels. 



As one approaches a quantum critical point from the insulating phase, the upper bound in Eq.~\eqref{eq:_TrS2_bounds} may diverge due to the closing of the energy gap. In the case where the critical point is conformal, the quantum metric also diverges and follows this universal behavior
\begin{flalign}
\frac{\Tr (\mathsf{S}_{2}^{ab})}{2}=\frac{\pi C_{J}}{2}\log(\xi).
\label{eq:_S2_CFT}
\end{flalign}
where $\xi$ is the correlation length of the system, and the universal coefficient $C_{J}$ is the current central charge. Less singular terms are neglected here. This logarithmic divergence results from the rigid form of the static structure factor in CFTs~\cite{bootstrapRMP}. The special case of free Dirac fermions has already been discussed in Refs.~\cite{Fu2024-1,CI-NI2006}.
Note that this universal behavior does not strictly require conformal invariance; an example of this will be discussed in Sec.~\ref{subsec:_Latt_CFL}. The logarithmic divergence in Eq.~\eqref{eq:_S2_CFT} is consistent with the logarithmic corner term in Eq.~\eqref{eq:_BF_CFT}. Quantum geometry clearly provides a unified framework for understanding previous results~\cite{Dirac_log,wulog,uni_corner,chengdisop,BF_CFS}, with Eq.~\eqref{eq:_BF_CFT} and Eq.~\eqref{eq:_BF_QH} as special cases.





\section{Harmonic Expansion}
\label{sec:_Dis_Sym}


In this section, we investigate the universal and non-universal features of corner charge fluctuations in both continuous and lattice systems. For continuous systems, we show that the orientation-resolved corner term $\gamma(\phi,\theta)$ can be expressed as an infinite series, as shown in Eq.~\eqref{eq:_BF_corner_expd}. This series consists of generalized universal angle functions defined in Eq.~\eqref{eq:_angle_functions}, with coefficients Eq.~\eqref{eq:_angle_functions_coef} determined by the harmonic expansion of the static structure factor in Eq.~\eqref{eq:_SSF_2}. For lattice systems, we present discrete formulas that apply under specific commensurability conditions and are expected to recover the continuous formulas in the long-wavelength limit, provided that the system's correlation length is sufficiently large. For both continuous and lattice models, we show that the small-$\theta$ limit is fully governed by the quantum metric, as described in Eq.~\eqref{eq:_BF_small-angle}.

\subsection{Continuous Systems}

\label{subsec:_Fourier}

In this subsection, we consider a continuous system that may arise from the long-wavelength description of certain lattice models. The inversion-symmetric part  $S_{+}(\boldsymbol{r})$ of the static structure factor generally has the following harmonic expansion
\begin{flalign}
S_{+}(\boldsymbol{r})=\sum_{n\in\mathbb{Z}}S_{n}(r)e^{\mathtt{i}nN\varphi},
\label{eq:_SSF_2}
\end{flalign}
where $\boldsymbol{r}=r(\cos\varphi,\sin\varphi)$, and the radial functions $S_{n}(r)$ are complex-valued. We introduce the integer $N$ to account for systems with $C_{N}$ rotational symmetry. Even in the absence of discrete rotational symmetry, the Fourier series in Eq.~\eqref{eq:_SSF_2} remains applicable with $N=2$. Following the procedure introduced in Ref.~\cite{uni_corner} (also see App.~\ref{app:_Fourier}), one can get rid of the boundary law and isolate the corner term by 
\begin{flalign}
\gamma(\phi,\theta)=-\int_{\textrm{B}}\textrm{d}^{2}\boldsymbol{r}_{1}\int_{\textrm{D}}\textrm{d}^{2}\boldsymbol{r}_{2}\,S_{+}(\boldsymbol{r}_{1}-\boldsymbol{r}_{2}),
\label{eq:_BF_corner}
\end{flalign}
where B and D are the regions introduced in FIG.~\ref{fig:_corner}. After substituting Eq.~\eqref{eq:_SSF_2} into Eq.~\eqref{eq:_BF_corner}, the corner contribution can be expanded as follows
\begin{flalign}
\gamma(\phi,\theta)=\sum_{n\in\mathbb{Z}}\mathsf{C}_{n}f_{n}(\theta)e^{\mathtt{i}nN\phi}.
\label{eq:_BF_corner_expd}
\end{flalign}
Here, we introduce a series of generalized {\it universal angle functions} $f_{n}(\theta)$ which capture the shape dependence of the $nN$-th harmonics of the orientation angle $\phi$ on the corner angle $\theta$. These functions have the following general expressions, as demonstrated in App.~\ref{app:_Fourier}
\begin{widetext}
\begin{flalign}
f_{n}(\theta)=\frac{4}{nN(n^2N^2-4)}\Big[2\cot(\theta)\sin\Big(\frac{nN}{2}(\theta-\pi)\Big)-nN\cos\Big(\frac{nN}{2}(\theta-\pi)\Big)\Big]\exp\Big(\mathtt{i}\frac{nN}{2}(\theta+\pi)\Big).
\label{eq:_angle_functions}
\end{flalign}
\end{widetext}
The complex conjugate satisfies $f_{n}^{*}(\theta)=f_{-n}(\theta)$. Notice that for $nN=0$ and $nN=\pm2$ these expressions should be understood through a limiting procedure, i.e., 
\begin{flalign}
f_{n}(\theta)=\begin{cases}
1+(\pi-\theta)\cot(\theta) & nN=0\\
-\frac{1}{2}(\cos(\theta)+\frac{\pi-\theta}{\sin(\theta)})e^{\pm\mathtt{i}\theta} & nN=\pm2
\end{cases}.
\end{flalign}
Therefore, $\lim_{n\rightarrow0}f_{n}(\theta)$ exactly reproduces the previously termed ``super-universal corner function''~\cite{uni_corner}, as given in Eq.~\eqref{eq:_angle_function_0}. The complex-valued {\it non-universal coefficients} $\mathsf{C}_{n}$ in Eq.~\eqref{eq:_BF_corner_expd} are given by
\begin{flalign}
\mathsf{C}_{n}=\int_{0}^{+\infty}\textrm{d}r\,\frac{-r^{3}}{2}S_{n}(r).
\label{eq:_angle_functions_coef}
\end{flalign}
where the radial functions $S_{n}(r)$ are defined in Eq.~\eqref{eq:_SSF_2}. It is clear that $\mathsf{C}_{0}$ is a real number, and generally, $\mathsf{C}_{n}^{\ast}=\mathsf{C}_{-n}$ when $|n|\geq1$.

There are several other notable features of  Eq.~\eqref{eq:_angle_functions}. Firstly, all angle functions vanish at $\theta=\pi$ and exhibit the following expansion around $\theta\approx\pi$
\begin{flalign}
f_{n}(\theta)\approx\frac{(-1)^{nN}}{3}(\theta-\pi)^{2}.
\end{flalign}
Moreover, Eq.~\eqref{eq:_angle_functions} 
 can show a singularity of $1/\theta$ as $\theta\rightarrow0$. For  $nN\neq0,\pm2$, the angle functions behave as 
\begin{flalign}
f_{n}(\theta)\approx\mathtt{i}\frac{4((-1)^{nN}-1)}{nN(n^{2}N^{2}-4)}\frac{1}{\theta}.
\label{eq:_angle_functions_singularity_1}
\end{flalign}
This expression vanishes when $nN$ is even. Due to the symmetry property $S_+(\boldsymbol{r})=S_+(-\boldsymbol{r})$, even in systems with $C_3$ symmetry, the structure factor $S_+$ effectively has $N=6$. Therefore, we can always assume $N$ is even. The $1/\theta$ singularity appears only when $nN$ equals $0$ or $\pm2$. The small-$\theta$ expressions in these cases are given by
\begin{flalign}
f_{n}(\theta)\approx\begin{cases}
+\pi/\theta & nN=0\\
-\pi/(2\theta) & nN=\pm2
\end{cases}.
\label{eq:_angle_functions_singularity_2}
\end{flalign}
Furthermore, while the real part of Eq.~\eqref{eq:_angle_functions} satisfies $\textrm{Re}[f_{n}(\theta)]=\textrm{Re}[f_{n}(2\pi-\theta)]$, the imaginary parts of $f_{n}(\theta)$ and $f_{n}(2\pi-\theta)$ are generally different when $n\neq0$.

The coefficients $\mathsf{C}_{n}$ of a few leading-order angle functions in Eq.~\eqref{eq:_BF_corner_expd} are entirely determined by quantum geometry. As demonstrated in App.~\ref{app:_Fourier}, the zeroth term for any $N\geq1$ satisfies
\begin{flalign}
\mathsf{C}_{0}=\frac{\Tr(\mathsf{S}_{2}^{ab})}{2\pi}.
\label{eq:_angle_function_0_coef}
\end{flalign}
For the case of $N=2$ and $n=\pm1$, we find that 
\begin{flalign}
\mathsf{C}_{\pm1}=\frac{\mathsf{S}_{2}^{xx}-\mathsf{S}_{2}^{yy}\mp\mathtt{i}2\mathsf{S}_{2}^{xy}}{2\pi}.
\label{eq:_angle_function_2_coef}
\end{flalign}
When $nN>2$, the coefficients $\mathsf{C}_{n}$ become more complicated and will be discussed in App.~\ref{app:_Fourier}.

It is noteworthy that the coefficients of terms exhibiting small-angle singularities, as shown in Eq.~\eqref{eq:_angle_functions_singularity_2}, are fully determined by the quantum metric $\mathsf{S}_{2}^{ab}$, as described by Eq.~\eqref{eq:_angle_function_0_coef} and Eq.~\eqref{eq:_angle_function_2_coef}. We can combine these terms to find the total singular contribution. For instance, for a system with $C_{2}$ symmetry, we have
\begin{flalign}
\gamma(\phi,\theta\rightarrow0)\approx\frac{\pi}{\theta}\mathsf{C}_{0}-\frac{\pi}{2\theta}(\mathsf{C}_{+1}e^{+\mathtt{i}2\phi}+\mathsf{C}_{-1}e^{-\mathtt{i}2\phi}),
\end{flalign}
which is equivalent to Eq.~\eqref{eq:_BF_small-angle}. In systems with $C_3$, $C_4$, or $C_6$ rotational symmetry, where the quantum metric is isotropic (i.e., $\mathsf{S}_{2}^{ab}=\mathsf{S}_{2}\delta^{ab}$), only the zeroth term in Eq.~\eqref{eq:_angle_functions_singularity_2} and Eq.~\eqref{eq:_angle_function_0_coef} contribute. This results in 
\begin{equation}
  \gamma(\phi,\theta\rightarrow0)\approx\frac{\Tr\mathsf{S}_{2}^{ab}}{2\theta}=\frac{\mathsf{S}_{2}}{\theta} . 
\end{equation}


The orientation-averaged bipartite fluctuations defined in Eq.~\eqref{eq:_BF_avg_def} turn out to be given by the simple formula Eq.~\eqref{eq:_BF_avg}, due to $\int_{0}^{2\pi}\textrm{d}\phi f_{n}(\theta)e^{\mathtt{i}nN\phi}=0$ for $n\neq0$. The small-angle behavior described in Eq.~\eqref{eq:_BF_avg_small-angle} can be easily verified using Eq.~\eqref{eq:_BF_avg}.

\subsection{On Lattice Models}
\label{subsec:_Latt}

\begin{figure}
\includegraphics[width=0.45\textwidth]{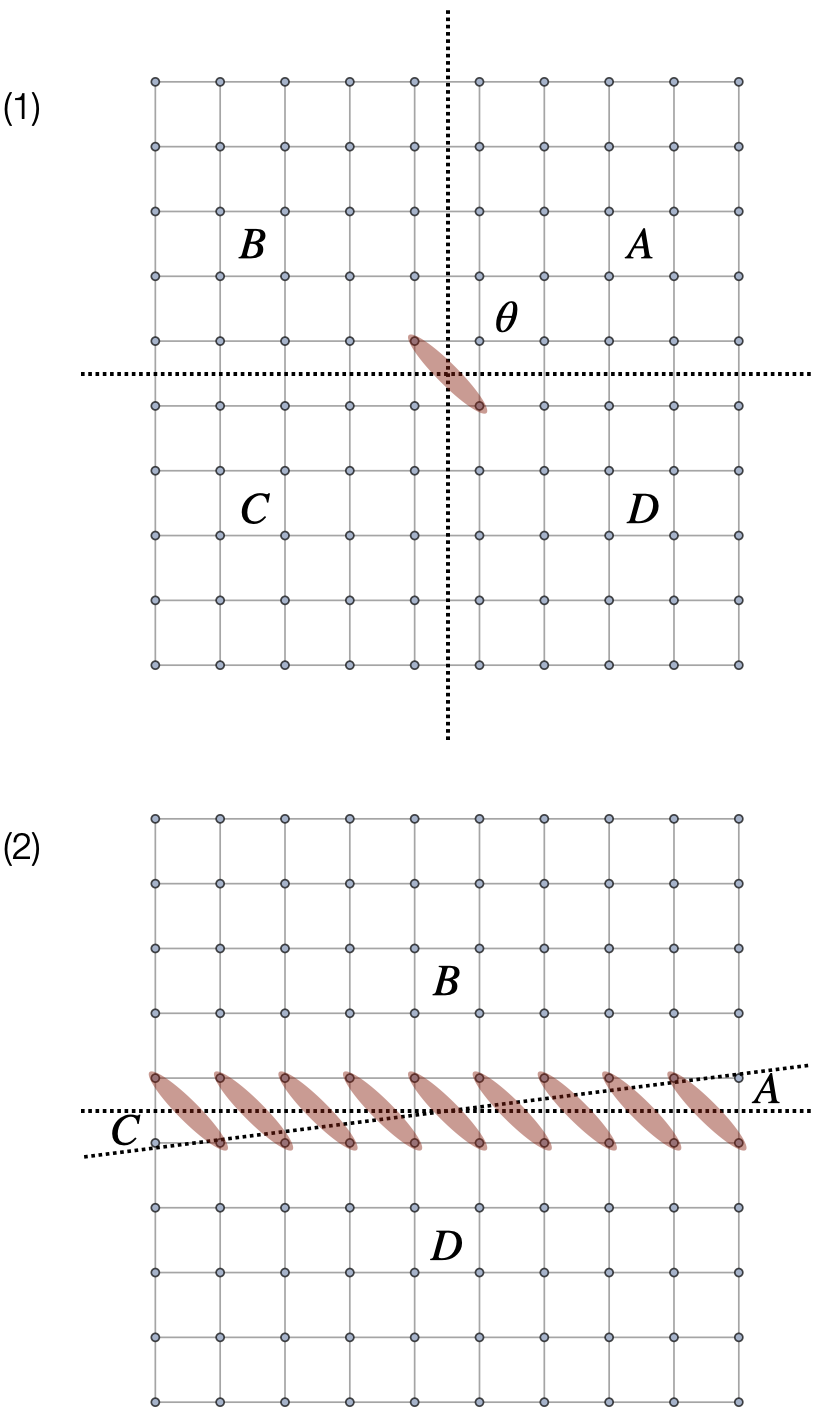}
\caption{The two configurations for square-lattice models with $C_{4}$ rotational symmetry. {\it (1)} When $\theta$ takes nonzero commensurate values, the number of bonds contributing to Eq.~\eqref{eq:_BF_corner_latt_BD} is given by Eq.~\eqref{eq:_bonds_number_1}. {\it (2)} In the small-$\theta$ limit, the number of bonds that contribute to Eq.~\eqref{eq:_BF_corner_latt_BD} is given by Eq.~\eqref{eq:_bonds_number_2}.}
\label{fig:_square_lattice}
\end{figure}

In this subsection, we examine lattice models more carefully without taking the long-wavelength limit. We still use $S_{+}(\boldsymbol{r})$ to denote the symmetrized equal-time density-density correlation, where $\boldsymbol{r}$ takes discrete values on the lattice. In the following, we assume that the lower edge of subsystem $\textrm{A}$ (i.e., the direction with $\phi=0$) is always aligned with one of the primitive vectors of the Bravais lattice.

For the finite subsystems depicted in FIG.~\ref{fig:_corner}, one can extract the summation of the following two corner terms
\begin{flalign}
\gamma(\phi,\theta)&=-\sum_{\boldsymbol{r}_{1}\in B}\sum_{\boldsymbol{r}_{2}\in D}S_{+}(\boldsymbol{r}_{1}-\boldsymbol{r}_{2}),\nonumber\\\gamma(\phi+\theta,\pi-\theta)&=-\sum_{\boldsymbol{r}_{1}\in C}\sum_{\boldsymbol{r}_{2}\in A}S_{+}(\boldsymbol{r}_{1}-\boldsymbol{r}_{2}),
\label{eq:_BF_corners_latt}
\end{flalign}
by $-(\mathcal{N}_{\textrm{A}}^{[2]}+\mathcal{N}_{\textrm{B}}^{[2]}+\mathcal{N}_{\textrm{C}}^{[2]}+\mathcal{N}_{\textrm{D}}^{[2]}-\mathcal{N}_{\textrm{AB}}^{[2]}-\mathcal{N}_{\textrm{CD}}^{[2]}-\mathcal{N}_{\textrm{BC}}^{[2]}-\mathcal{N}_{\textrm{AD}}^{[2]}+\mathcal{N}_{\textrm{ABCD}}^{[2]})/2$, where the leading-order boundary-law terms cancel out~\cite{corner_QM}. The first corner term $\gamma(\phi,\theta)$ can be further simplified to
\begin{flalign}
\sum_{\boldsymbol{r}_{1}\in B}\sum_{\boldsymbol{r}_{2}\in D}S_{+}(\boldsymbol{r}_{1}-\boldsymbol{r}_{2})=\sum_{\Delta\boldsymbol{r}}\mathbbm{n}(\Delta\boldsymbol{r})S_{+}(\Delta\boldsymbol{r}),
\label{eq:_BF_corner_latt_BD}
\end{flalign}
where $\mathbbm{n}(\Delta\boldsymbol{r})$ represents the number of bonds connecting sites in B and D with a separation of $\Delta\boldsymbol{r}=\boldsymbol{r}_{1}-\boldsymbol{r}_{2}$. The second corner term can be determined similarly.

For systems with $C_{N}$ symmetry, we align the origin with the symmetry axis of the rotations when cutting the subsystems. At commensurate angles $0<\theta<\pi$ and $0<\phi<2\pi$ (i.e.,  integer multiples of $2\pi/N$), we find that
\begin{flalign}
\mathbbm{n}(\Delta\boldsymbol{r})=\frac{\Delta r^{a}\Delta r^{b}}{V_{\textrm{cell}}}\Lambda_{ab}(\phi,\theta),
\label{eq:_bonds_number_1}
\end{flalign}
where $V_{\textrm{cell}}$ is the volume of the unit cell and 
\begin{flalign}
\Lambda_{ab}(\phi,\theta)=\frac{1}{\sin\theta}\left(\begin{array}{cc}
\sin\phi\sin(\theta+\phi) & -\frac{1}{2}\sin(\theta+2\phi)\\
-\frac{1}{2}\sin(\theta+2\phi) & \cos\phi\cos(\theta+\phi)
\end{array}\right).
\end{flalign}
One example on the square lattice is illustrated in example (1) in FIG.~\ref{fig:_square_lattice}. The number of bounds $\mathbbm{n}(\Delta\boldsymbol{r})$ for $\boldsymbol{r}_{1}\in C$ and $\boldsymbol{r}_{2}\in A$ can be easily determined by noting that $\Lambda_{ab}(\phi+\theta,\pi-\theta)=-\Lambda_{ab}(\phi,\theta)$. In principle, the exact result for Eq.~\eqref{eq:_BF_corners_latt} can be determined for lattice models on a case-by-case basis using the provided formulas. To relate these results to the long-wavelength results discussed in Sec.~\ref{subsec:_Fourier}, it is necessary to ensure that the subsystem size and the correlation length are significantly larger than the lattice constant.

In addition to analyzing finite commensurate values of $
\theta$, we can also gain insights into the small-$
\theta$ limit. When $\Delta\boldsymbol{r}$ is small, the value of $\mathbbm{n}(\Delta\boldsymbol{r})$ in Eq.~\eqref{eq:_BF_corner_latt_BD} is insensitive to small variations in $\theta$ when $\theta$ is relatively large. In contrast, in the $\theta\rightarrow0$ limit, the number of bonds becomes very large even for a short separation $\Delta\boldsymbol{r}$, as shown in example (2) in  FIG.~\ref{fig:_square_lattice}. In the thermodynamic limit, this can be estimated as 
\begin{flalign}
\mathbbm{n}(\Delta\boldsymbol{r})\approx\frac{1}{V_{\textrm{cell}}}\frac{(\Delta\boldsymbol{r}\cdot\hat{\tau}(\phi))^{2}}{\theta},
\label{eq:_bonds_number_2}
\end{flalign}
where $\hat{\tau}(\phi)=(-\sin\phi,\cos\phi)$ is the unit vector perpendicular to the lower edge of region A. Another special property of the $\theta\rightarrow0$ limit is that $\Delta\boldsymbol{r}$ is allowed to take half of all possible values on the lattice, with its direction lying between $0$ and $\pi$. Thanks to the symmetry $S_{+}(\Delta\boldsymbol{r})=S_{+}(-\Delta\boldsymbol{r})$, the summation in Eq.~\eqref{eq:_BF_corner_latt_BD} can be extended to the entire lattice. Thus, we find that
\begin{flalign}
&\gamma(\phi,\theta\rightarrow0)=\frac{1}{2}\frac{1}{\theta}\lim_{\boldsymbol{q}\rightarrow0}(\hat{\tau}(\phi)\cdot\partial_{\boldsymbol{q}})^{2}S_{+}(\boldsymbol{q})\nonumber\\=\;&\frac{1}{\theta}(\sin^{2}(\phi)\mathsf{S}_{2}^{xx}+\cos^{2}(\phi)\mathsf{S}_{2}^{yy}-\sin(2\phi)\mathsf{S}_{2}^{xy}),
\label{eq:_BF_small-angle_Latt}
\end{flalign}
where $S_{+}(\boldsymbol{q})$ is obtained from the Fourier transformation of $S_{+}(\boldsymbol{r})$ and follows the same convention as Eq.~\eqref{eq:_SSF_1}. In other words, we have verified the general formula Eq.~\eqref{eq:_BF_small-angle} in lattice systems. The result in Eq.~\eqref{eq:_BF_small-angle_Latt} has recently appeared in Ref.~\cite{corner_QM} in the context of the band topology and geometry of free fermions. Here, we emphasize that it also applies to strongly correlated systems, provided the quantum metric is properly defined using twisted boundary conditions.

\section{Chern Insulators\label{sec:_Chern_insulators}}



In this section, we consider the quantum metric $\mathsf{S}_2^{ab}$ and corner contribution in lattice realization of Chern insulator. We start with a general two-band Hamiltonian in two spatial dimensions
\begin{align}
  H = \sum_{\boldsymbol{k}} \begin{pmatrix}
  c_{A,\boldsymbol{k}}^{\dagger} & c_{B,\boldsymbol{k}}^{\dagger}
  \end{pmatrix}
  \mathcal{H}(\boldsymbol{k})
  \begin{pmatrix}
  c_{A,\boldsymbol{k}} \\
  c_{B,\boldsymbol{k}}
  \end{pmatrix},
\end{align}
where $\mathcal{H}(\boldsymbol{k})=\boldsymbol{d}(\boldsymbol{k})\cdot\boldsymbol{\sigma}$ and $\boldsymbol{\sigma}$ is the vector of Pauli matrices.
The static structure factor $S(\boldsymbol{q})$ can be determined after explicitly diagonalizing the Hamiltonian. We have
\begin{align}
  S(\boldsymbol{q})&=\frac12 \int_{\text{BZ}}\frac{\textrm{d}^{2}\boldsymbol{k}}{(2\pi)^2} \left( 1-\hat{\boldsymbol{d}}(\boldsymbol{k})\cdot\hat{\boldsymbol{d}}(\boldsymbol{k}+\boldsymbol{q})\right).
\end{align}
Here $\hat{\boldsymbol{d}}=\boldsymbol{d}/|\boldsymbol{d}|$ represents the normalized vector.
For $\boldsymbol{q}$ small, we can expand $S(\boldsymbol{q})$ up to second order as 
\begin{align}
  S(\boldsymbol{q})=\frac{q_{a}q_{b}}{4}\int_{\textrm{BZ}}\frac{\textrm{d}^{2}\boldsymbol{k}}{(2\pi)^{2}}\frac{\partial\hat{\boldsymbol{d}}(\boldsymbol{k})}{\partial k_{a}}\cdot\frac{\partial\hat{\boldsymbol{d}}(\boldsymbol{k})}{\partial k_{b}}+\mathscr{O}(|\boldsymbol{q}|^{4}).
\end{align}

In the presence of $C_3$ or $C_4$ rotational symmetry, the quantum metric is proportional to the identity, i.e., $\mathsf{S}_{2}^{ab}=\mathsf{S}_2 \delta^{ab}$, and the static structure factor simplifies to
\begin{align}
  S(\boldsymbol{q})&=\frac18 q^2 \int_{\text{BZ}}\frac{\textrm{d}^2\boldsymbol{k}}{(2\pi)^2} (\nabla\hat{\boldsymbol{d}})^2+\mathscr{O}(|\boldsymbol{q}|^4)\nonumber\\
  &=\mathsf{S}_2 q^2+\mathscr{O}(|\boldsymbol{q}|^4).\label{Eq_S2SquareLatticeHaldane}
\end{align}
In Eq.~\eqref{Eq_S2SquareLatticeHaldane}, $\mathsf{S}_2$ exactly matches the Hamiltonian of an O(3) sigma model defined in momentum space. The Chern number $C$ translates to the number of topological charges in the system. And the topological lower bound on $\mathsf{S}_2$ follows from the Bogomoln’yi inequality~\cite{bogomol1976stability, belavin1975pseudoparticle}, which gives the lower bound on energy in the presence of topological charges
\begin{align}
    E=\frac12 \int\textrm{d}^2\boldsymbol{k} (\nabla\hat{\boldsymbol{d}})^2 &\geq \pm\frac{1}{2} \int\textrm{d}^2\boldsymbol{k} \epsilon^{\mu\nu}\hat{\boldsymbol{d}}\cdot(\partial_{\mu}\hat{\boldsymbol{d}}\times\partial_{\nu}\hat{\boldsymbol{d}})\notag\\
    &=\pm 4\pi C.\label{Eq_BPineq}
\end{align}

To achieve equality in the bound, the integrands must be equal at every point.
It is helpful to introduce $\mathbb{C}P^1$ coordinate $W(z,\bar{z})=\frac{\hat{d}_1+\mathtt{i}\hat{d}_2}{1-\hat{d}_3}$ with $z=k_x+\mathtt{i}k_y$. Then we have
\begin{align}
    \mathsf{S}_2=\frac{1}{16\pi^2}\int\frac{|\partial_z W|^2+|\partial_{\bar{z}} W|^2}{(1+|W|^2)^2}\textrm{d}z\textrm{d}\bar{z},
\end{align}
and
\begin{align}
    C=\frac{1}{4\pi}\int\frac{|\partial_z W|^2-|\partial_{\bar{z}} W|^2}{(1+|W|^2)^2}\textrm{d}z\textrm{d}\bar{z}.
\end{align}
In the following discussion, we will focus on the case where $C>0$. $\mathsf{S}_2$ is minimized when $\partial_{\bar{z}}W\equiv 0$, which means $W=W(z)$ is a meromorphic function. Meanwhile, such a function automatically ensures that the off-diagonal terms in $\mathsf{S}_{2}^{ab}$ vanish, as
\begin{align}
    \partial_{k_x}\hat{\boldsymbol{d}}\cdot \partial_{k_y}\hat{\boldsymbol{d}}=4\mathtt{i}
    \frac{\partial_{\bar{z}}W\partial_{\bar{z}}\bar{W}-\partial_{z}W\partial_{z}\bar{W}}{(1+W\bar{W})^2}.
\end{align}

In the continuum, the simplest example is $W(z)=\lambda z$, where $\lambda$ is a constant real parameter. The corresponding model is
\begin{align}
    \hat{d}(\boldsymbol{k})=\left(\frac{2\lambda k_x}{1+\lambda^2 k^2},\frac{2\lambda k_y}{1+\lambda^2 k^2},\frac{1-\lambda^2 k^2}{1+\lambda^2 k^2}\right).
\end{align}
This includes the special case with a parabolic spectrum discovered in Ref.~\cite{Fu2024-3}, where
\begin{align}
    d(\boldsymbol{k})=\left(\Delta_p k_x,\Delta_p k_y,-\frac{k^2}{2m}+\frac{m\Delta_p^2}{2}\right).
\end{align}
In lattice models, an additional requirement is that $W(z)$ is doubly periodic, i.e. $W(z)=W(z+2\pi)=W(z+2\pi\mathtt{i})$. Consequently, $W(z)$ is a elliptic function, whose pole structure determines the Chern number $C$. Given that there are at least two poles in the Brillouin zone, the bound cannot be saturated for the $C=1$~\cite{jian2013momentum} case of the two-band Hamiltonian. We note that ideal topological bands recently studied in Refs. \cite{Wang2021,Liu2024} are known to saturate the topological bounds.

It is worth noting that the same structure appears in a different context in Ref.~\cite{jian2013momentum}, where they consider the maximally localized flat-band Hamiltonian. With the flat-band restriction $|\boldsymbol{d}|=1$, the ``mean hopping range'' they defined is proportional to $\mathsf{S}_2$ in Eq.~\eqref{Eq_S2SquareLatticeHaldane}, which then has a topological lower bound for the same reason here.

\begin{figure}[t]
  \centering
  \includegraphics[width=\linewidth]{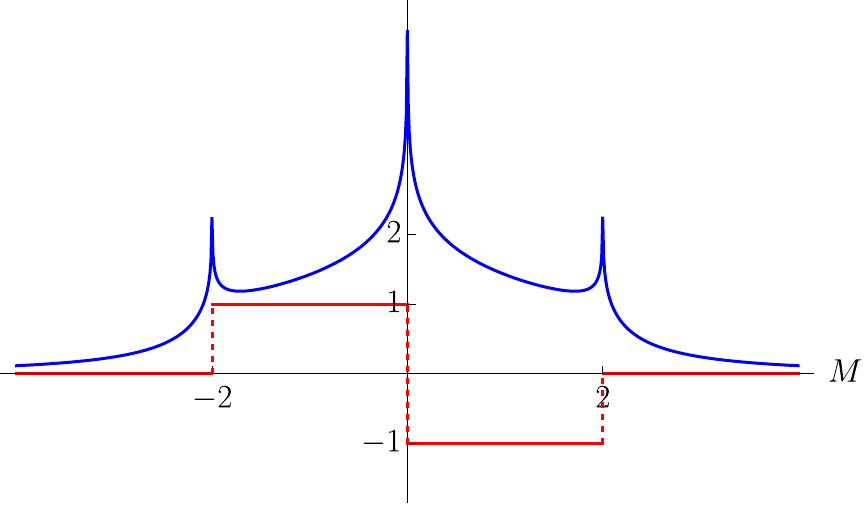}
  \caption{Comparison of the quantum metric $4\pi \mathsf{S}_2$ (blue line) and the Chern Number (red line) for the square-lattice Haldane Model. The quantum metric diverges logarithmically near the topological transition.}
  \label{Fig_HaldaneModelS2}
\end{figure}

In FIG.~\ref{Fig_HaldaneModelS2}, we plot numerical results of $\mathsf{S}_2$ in the square-lattice Haldane model, where
\begin{align}
  \boldsymbol{d}(\boldsymbol{k})=(\sin k_x,\sin k_y,M-\cos k_x-\cos k_y).
\end{align}
The bound is not saturated over the entire parameter range. Moreover, near the topological transition, $S_2$ diverges logarithmically. This divergence can be understood from the long-wavelength continuous theory with
\begin{align}
    \boldsymbol{d}(\boldsymbol{k})=(k_x,k_y,M-2),
\end{align}
where we can show that $\mathsf{S}_{2}\approx\frac{1}{16\pi}\log\frac{1}{|M-2|}$ near the critical point $M=2$.

\begin{figure}[t]
  \centering
  \includegraphics[width=\linewidth]{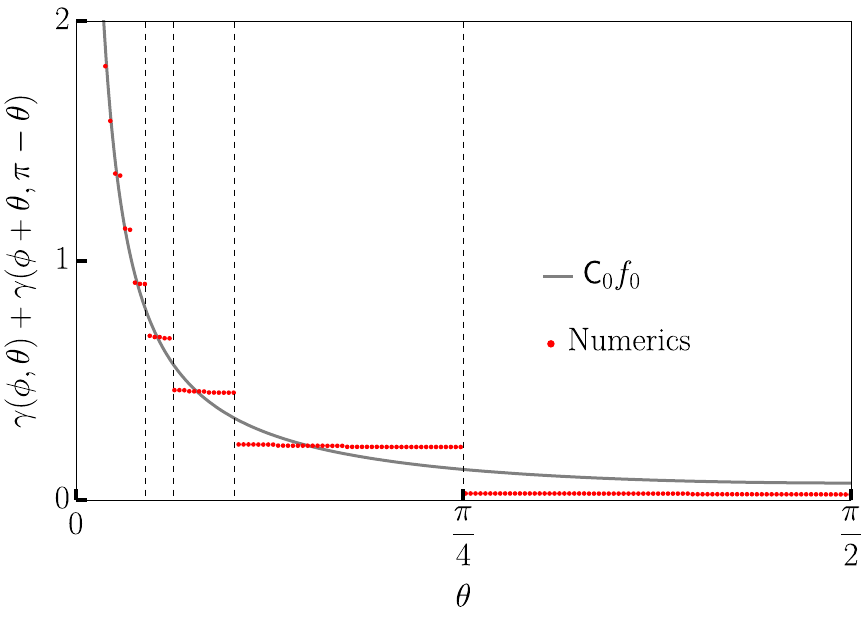}
  \caption{The corner term for $\phi=0$ and $M=1$. Numerical results are plotted using red dots, while the analytical prediction from the quantum metric is shown by the gray line. The origin is aligned the symmetry axis of the rotations when cutting the subsystems. Jumps in the numerical results correspond to special values of $\theta$ where the sites in the subsystems change drastically with small variations in $\theta$. Here, they are determined by $\theta=\arctan(1/n)$, where $n=1,3,5,\cdots$.}
  \label{Fig_HaldaneModelCorner1}
  \includegraphics[width=\linewidth]{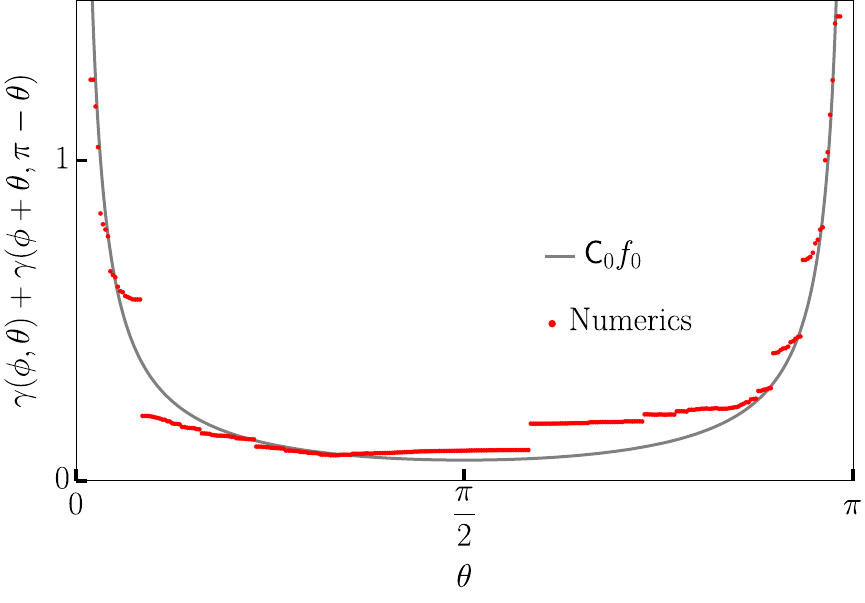}
  \caption{The corner term for $\phi=\pi/6$ and $M=1.9$. For $\phi\neq 0$, the corner term from lattice calculations are no longer symmetric in general, and the jumps are also shifted.}
  \label{Fig_HaldaneModelCorner2}
\end{figure}

Now we compare the corner term $\gamma(\phi,\theta)$ extracted numerically in the square-lattice Haldane model to the results of continuous system with $C_N$ rotational symmetry in Sec.~\ref{subsec:_Fourier}. Here $N=4$, and later we will also consider $N=2$ by adding anisotropic hoppings. In our numerical calculations the total system size is $50\times 50$, and the size of the subsystem A is roughly 25. We use the strategy discussed in Sec.~\ref{subsec:_Latt} to extract the summation of two corner contributions in Eq.~\eqref{eq:_BF_corners_latt}:
\begin{align}
    \gamma(\phi,\theta)+\gamma(\phi+\theta,\pi-\theta).
\end{align}

The $\phi=0$ and $M=1$ case is plotted in FIG.~\ref{Fig_HaldaneModelCorner1}. Here, we only consider $\theta\in[0,\pi/2]$, since the summation of two corner terms is symmetric with respect to $\theta=\pi/2$. When $\theta\rightarrow 0$, $\gamma(\phi,\theta)$ is accurately represented by $\mathsf{C}_0 f_0\approx \Tr \mathsf{S}_2^{ab}/(2\theta)$, consistent with our earlier discussion. For larger $\theta$, the regions defined on the lattice is not sensitive to small changes in $\theta$ except at certain special values, resulting in plateaus and discontinuities in the $\theta$ dependence of the $\gamma(\phi,\theta)$. On average, the corner term still follows the form determined by the quantum metric.

\begin{figure}[t]
  \centering
  \includegraphics[width=\linewidth]{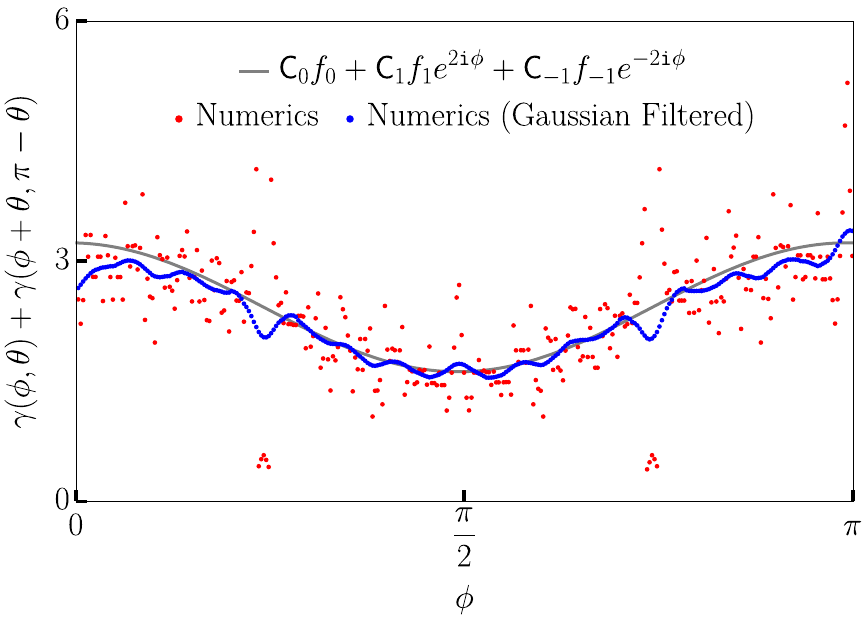}
  \caption{The corner term for $t_x=0.8$, $t_y=1.2$, $M=1$ and $\theta=0.05$. In this figure, we add the gaussian filtered numerical data with standard deviation $\sigma=5$ in blue, which shows that the $\phi$ dependence of $\gamma$ is described by our analytical formula on average.}
  \label{Fig_HaldaneModelCorner3}
\end{figure}

For general values of $\phi$, the summation of two corner terms is not symmetric with respect to $\theta=\pi/2$. And the discontinuities of lattice results are shifted. However, the $\gamma(\phi,\theta)$ still agrees well with our analytical result on an average sense. It is also worth noting that, as we tune the parameter $M$ closer to the topological transition, the approximation near $\theta \sim \pi/2$ improves. This is because the correlation length and $S_2$ diverges. We plot the $\phi=\pi/6$ and $M=1.9$ case in FIG.~\ref{Fig_HaldaneModelCorner2} for a demonstration.

We can also reduce the $C_4$ symmetry of the square-lattice Haldane model to $C_2$ by introducing different hopping amplitudes along different axes:
\begin{align}
  \boldsymbol{d}(\boldsymbol{k})=(t_x \sin k_x,t_y \sin k_y,M-t_x\cos k_x-t_y\cos k_y).
\end{align}
For $t_x\neq t_y$, the quantum metric $\mathsf{S}_2^{ab}$ is still diagonal, but $\mathsf{S}_2^{xx}\neq\mathsf{S}_2^{yy}$. The leading contributions to $\gamma(\phi,\theta)$ are $\mathsf{C}_{0}f_{0}+\mathsf{C}_{1}f_{1} e^{2\mathtt{i}\phi} +\mathsf{C}_{-1}f_{-1} e^{-2\mathtt{i}\phi}$ where $\mathsf{C}_{\pm1}=(\mathsf{S}_{2}^{xx}-\mathsf{S}_{2}^{yy})/(2\pi)$. In this case, the quantum metric causes non-trivial $\phi$ dependence of the corner term through $f_{\pm 1}$. We plot $\gamma(\phi,\theta)$ as a function of $\phi$ for $\theta=0.05$ in Fig. \ref{Fig_HaldaneModelCorner3}. It is clearly seen that the data points of $\gamma(\phi,\theta)$ fluctuates significantly with $\phi$, most likely due to lattice effects, although the overall trend follows the analytical result. We also perform Gaussian filtering on the data, which now shows a reasonable agreement with the formula.


\section{Anisotropic Landau Levels}
\label{sec:_Ani_LL}

In this section, we discuss several general results regarding the (anisotropic) Landau levels of non-relativistic particles. The original Kohn theorem~\cite{Kohn1964} applies to systems that are invariant under the full Galilean group, which includes continuous translations, Galilean boosts, and continuous rotations. However, we emphasize that Galilean boosts and translations alone are sufficient to establish a generalized Kohn theorem, which we prove in App.~\ref{app:_Kohn_th}. In Sec.~\ref{subsec:_Kohn}, we describe how the theorem ensures that the many-body quantum metric $\mathsf{S}_{2}^{ab}$ is determined by UV information, with a particular focus on its implications for corner charge fluctuations.

Generalized rotation invariance, defined with respect to the quantum metric, allows us to derive an exact analytical expression for the orientation-resolved corner term in Eq.~\eqref{eq:_BF_corner} without the need to sum the infinite series in Eq.~\eqref{eq:_BF_corner_expd}. We discuss its implications in Sec.~\ref{subsec:_QM_Sym}, focusing on the special case where the conditions for the generalized Kohn theorem are also fulfilled. In this scenario, the bipartite fluctuations are given by Eq.~\eqref{eq:_BF_QH_2}. In the isotropic limit, where the system becomes invariant under the Galilean group, it is evident that this result reduces to the previous formula Eq.~\eqref{eq:_BF_QH} in the literature~\cite{uni_corner,BF_CFS}. Further details on the most general case of generalized rotation invariance are provided in App.~\ref{app:_QM_Sym}.



In Sec.~\ref{subsec:_FQHE}, we investigate the fractional QH states in anisotropic LLs in greater detail. Using a generating function approach, we discuss several classes of examples where $\mathsf{S}_2^{ab}$ can be computed analytically.  
Holomorphicity of the many-body wavefunction in the center of mass coordinate is found to be a general condition that leads to the saturation of the bound on $\mathsf{S}_2^{ab}$. We also illustrate a set of examples where the bound is not saturated.


\subsection{Generalized Kohn Theorem \label{subsec:_Kohn}}

We consider a system of $N_{\textrm{e}}$ non-relativistic particles, each with unit electric charge and labeled by $j=1,2,\ldots, N_{\textrm{e}}$, in a uniform perpendicular magnetic field $B$. The many-body Hamiltonian is given by
\begin{equation}
H=\sum_{j=1}^{N_{\textrm{e}}}\frac{\eta^{ab}}{2m}\pi_{a}^{j}\pi_{b}^{j}+\sum_{j\neq i}U(\boldsymbol{r}_{i}-\boldsymbol{r}_{j}).
\label{eq:_Kohn_Hamiltonian}
\end{equation}
Here, $\pi_{a}^{j}=p_{a}^{j}-A_{a}(\boldsymbol{r}_{j})$
denotes the gauge-invariant momentum where $\boldsymbol{A}$ is the background field such that $\nabla \times \bm A= B$. The  mass tensor is defined as $m\eta_{ab}$, where $\eta_{ab}$ is an unimodular metric satisfying $\det(\eta_{ab})=1$, and $m>0$ controls the ``size'' of the mass. The mass tensor is generally anisotropic and positive-definite. In Eq.~\eqref{eq:_Kohn_Hamiltonian}, the metric $\eta^{ab}$ is the inverse of $\eta_{ab}$ such that $\eta^{ab}\eta_{bc}=\delta_{c}^{a}$. We also include an inversion-symmetric interaction term $U(\boldsymbol{r})=U(-\boldsymbol{r})$. For the Coulomb potential at long distances, $U(\boldsymbol{r})$ is typically a function of the distance $\widetilde{\eta}_{ab}r^{a}r^{b}$, where $\widetilde{\eta}_{ab}$ is another unimodular metric that may be anisotropic due to the dielectric properties of the materials. In Ref.~\cite{haldane_geom}, $\eta_{ab}$ is referred to as the Galilean metric, while $\widetilde{\eta}_{ab}$ is called the Coulomb metric. 

The Hamiltonian Eq.~\eqref{eq:_Kohn_Hamiltonian} is invariant under Galilean boosts, and the center of mass degree of freedom can still be separated from the relative degrees of freedom. This can be seen explicitly through the following change of coordinates $\boldsymbol{r}_{j}^{\prime}=\bar{\boldsymbol{r}}$ for $j=N_{\textrm{e}}$ and $\boldsymbol{r}_{j}^{\prime}=\boldsymbol{r}_{j}-\bar{\boldsymbol{r}}$ for $1\leq j\leq N_{\textrm{e}}-1$, where $\bar{\boldsymbol{r}}=(1/N_{\textrm{e}})\sum_{j=1}^{N_{\textrm{e}}}\boldsymbol{r}_{j}$ is the center of mass coordinate. With this transformation, the Hamiltonian in Eq.~\eqref{eq:_Kohn_Hamiltonian} can be simplified to $H=H_{\textrm{CoM}}+H_{\rm Rel}$, where 
\begin{flalign}
H_{\textrm{CoM}}=\frac{\eta^{ab}\bar{\pi}_{a}\bar{\pi}_{b}}{2mN_{\textrm{e}}}
\label{eq:_Kohn_Hamiltonian_CoM}
\end{flalign}
depends only on the center of mass momentum $\bar{\boldsymbol{\pi}}$, which is conjugate to $\bar{\boldsymbol{r}}$. The relative part $H_{\rm Rel}$ is completely independent of $\bar{\boldsymbol{r}}$ and therefore commutes with $\bar{\boldsymbol{\pi}}$. {The decoupling of the center-of-mass degree of freedom results in the standard relation by Kohn~\cite{Kohn1964}, which states that the cyclotron frequency is not renormalized by interactions. Moreover, it constrains the quadratic term in the static structure factor Eq.~\eqref{eq:_SSF_1} to be} 
\begin{flalign}
\mathsf{S}_{2}^{ab}=\frac{|\nu|}{4\pi}\eta^{ab},
\label{eq:_S2_Kohn}
\end{flalign}
where $\nu = 2\pi N_e/(B V)$ is the filling factor, and $\eta^{ab}$ is the unimodular Galilean metric, characterizing the ``shape'' of the mass tensor in the UV. Importantly, $\mathsf{S}_{2}^{ab}$ is
independent of the ``size'' of the mass tensor represented by $m$, and the specific form of the interaction $U(\boldsymbol{r})$, provided that it depends only on the relative coordinates. We prove Eq.~\eqref{eq:_S2_Kohn} using two distinct approaches. In App.~\ref{app:_Kohn_th}, we show that it follows directly from response theory. Additionally, in Sec.~\ref{subsubsec:Galilean_invariance_saturation}, we will demonstrate that the same conclusion can be drawn from the holomorphic properties of ground-state wavefunctions. This is a general result for both compressible and incompressible QH states realized by the UV Hamiltonian in Eq.~\eqref{eq:_Kohn_Hamiltonian}. 


Based on the Hamiltonian Eq.~\eqref{eq:_Kohn_Hamiltonian}, and the separation between center of mass and relative degrees of freedom, the energetic upper bounds in Eq.~\eqref{eq:_TrS2_bounds}, Eq.~\eqref{eq:_detS2_bounds} and Eq.~\eqref{eq:_mTrS2_bounds} are given by the positive semi-definiteness 
\begin{equation}
\mathsf{S}_{2}^{ab}\leq\frac{1}{2\omega_{c}}\sD_2^{ab} =\frac{|\nu|}{4\pi}\eta^{ab},
\end{equation}
where the energy gap is given by the cyclotron frequency $\omega_{c}=1/(m\ell_{B}^{2})$. Therefore, the generalized Kohn theorem implies that the upper bound is always saturated. The topological lower bounds can be understood through the properties of the unimodular metric $\eta^{ab}$. We can see that the determinant bound Eq.~\eqref{eq:_detS2_bounds} is also saturated due to $\det(\eta^{ab})=1$,
while the trace bound in Eq.~\eqref{eq:_TrS2_bounds} is saturated only in the isotropic case $\eta^{ab}=\delta^{ab}$. This can be understood using the inequality of arithmetic and geometric means
\begin{widetext}
\begin{flalign}
\frac{\Tr(\mathsf{S}_{2}^{ab})}{2}=\frac{|\nu|}{4\pi}\frac{\eta^{xx}+\eta^{yy}}{2}\geq\frac{|\nu|}{4\pi}\sqrt{\eta^{xx}\eta^{yy}}=\frac{|\nu|}{4\pi}\sqrt{1+(\eta^{xy})^{2}}\geq\frac{|\nu|}{4\pi}.
\end{flalign}
\end{widetext}
Also, note that if the modified trace is defined using the Galilean metric $\eta_{ab}$, then the lower bound in Eq.~\eqref{eq:_mTrS2_bounds} is still formally satisfied. When relating these results to Eq.~\eqref{eq:_TrS2_bounds}, Eq.~\eqref{eq:_detS2_bounds} and Eq.~\eqref{eq:_mTrS2_bounds}, it is important to note that the filling factor is proportional to the Hall conductivity, i.e., $\sigma^{xy}=\nu /(2\pi)$, which is ensured by continuous translational symmetry, as demonstrated in App.~\ref{app:_translational_symmetry}.

The general formalism developed in Sec.~\ref{subsec:_Fourier} (and App.~\ref{app:_Fourier}) is applicable here. For systems without discrete rotational symmetry, we can still consider the Fourier transformation Eq.~\eqref{eq:_SSF_2} with $N=2$. After averaging over orientation, only the zeroth term from Eq.~\eqref{eq:_angle_functions} survives.  Moreover, Eq.~\eqref{eq:_angle_function_0_coef} remains valid in this case.
Therefore, we can make a general statement about the orientation-averaged corner term Eq.~\eqref{eq:_BF_avg_def} for anisotropic Landau levels
\begin{flalign}
\bar{\gamma}(\theta)=\frac{|\sigma^{xy}|}{2\pi}\frac{\Tr(\eta^{ab})}{2}(1+(\pi-\theta)\cot\theta),
\label{eq:_BF_QH_avg}
\end{flalign}
where $\Tr(\eta^{ab})\geq1$ is saturated only when $\eta^{ab}=\delta^{ab}$. 



\subsection{Generalized Rotational Invariance}
\label{subsec:_QM_Sym}


In this subsection, we assume that the system possesses a generalized rotational invariance, defined by the Hamiltonian commuting with the angular momentum operator
\begin{flalign}
L_{z}= \sum_{j=1}^{N_e}\frac{1}{2}\left(B\eta_{ab}R_{j}^{a}R_{j}^{b}-\frac{\eta^{ab}\pi_{a}^{j}\pi_{b}^{j}}{B}\right),
\label{eq:_ang_mom}
\end{flalign}
where $R_{j}^{a}=r_{j}^{a}+\epsilon^{ab}\pi_{b}^{j}/B$ is the guiding center coordinate. The nice property is that the system can be converted to an isotropic one through coordinate transformations.



We introduce a set of vielbeins $e_{I}^{a}$ to diagonalize the Galilean metric $\eta^{ab}=e_{I}^{a}e_{J}^{b}\delta^{IJ}$. The coordinate transformations of the various vectors are given by 
\begin{flalign}
\tilde{q}_{I}=e_{I}^{a}q_{a}\qquad\tilde{r}^{I}=e_{a}^{I}r^{a},\qquad\tilde{A}_{I}=e_{I}^{a}A_{a},
\label{eq:_coord_tran_UV}
\end{flalign}
where $e_{a}^{I}$ is the inverse of $e_{I}^{a}$, such that $e_{I}^{a}e_{b}^{I}=\delta_{b}^{a}$ and $e_{I}^{a}e_{a}^{J}=\delta_{I}^{J}$. We first consider the special case of Eq.~\eqref{eq:_Kohn_Hamiltonian} where the Galilean metric $\eta_{ab}$ and the Coulomb metric $\widetilde{\eta}_{ab}$ are identical. In the new isotropic system, the filling factor $\nu$ retains its original value\footnote{In the new coordinate system, the magnetic field strength is $\tilde{B}=\varepsilon^{IJ}\partial\tilde{A}_{J}/\partial\tilde{r}^{I}=B/\sqrt{\det\eta_{ab}}$, where $B=\varepsilon^{ab}\partial A_{b}/\partial r^{a}$ is the original magnetic field. We have used the fact that the Levi-Civita symbol is a tensor density of weight $-1$, which satisfies $\varepsilon^{ab}/\sqrt{\det\eta_{ab}}=\varepsilon^{IJ}e_{I}^{a}e_{J}^{b}$. Consequently, the filling factor $\nu=2\pi N/(\tilde{B}\tilde{V})$ remains invariant since the volume of the system transforms as $\tilde{V}=V\sqrt{\det\eta_{ab}}$.} and the static structure factor satisfies $\tilde{S}(\tilde{\boldsymbol{q}})=\frac{|\nu|}{4\pi}|\tilde{\boldsymbol{q}}|^{2}+\ldots$. The calculations in Ref.~\cite{uni_corner} are applicable here, and Eq.~\eqref{eq:_BF_QH} holds for the new system. The formalism for relating bipartite charge fluctuations between the old and new coordinate systems is developed in App.~\ref{app:_QM_Sym}. For the configuration shown in FIG.~\ref{fig:_corner}, the result is 
\begin{flalign}
\mathcal{N}_{\textrm{A}}^{[2]}(\phi,\theta)=\#L-\frac{|\sigma^{xy}|}{2\pi}f_{0}(\tilde{\theta}(\phi,\theta)).
\label{eq:_BF_QH_2}
\end{flalign}
The new angle variable $\tilde{\theta}$, defined for the subregion of the new isotropic system, can be determined from the original angles $\theta$ and $\phi$ using the relation
\begin{flalign}
\cos\tilde{\theta}=\frac{\eta_{ab}\hat{n}^{a}\hat{m}^{b}}{\sqrt{\eta_{ab}\hat{n}^{a}\hat{n}^{b}}\sqrt{\eta_{ab}\hat{m}^{a}\hat{m}^{b}}},
\end{flalign}
where $\hat{n}=(\cos\phi,\sin\phi)$ and $\hat{m}=(\cos(\phi+\theta),\sin(\phi+\theta))$. It is evident that the ``super-universal formula'' given in Eq.~\eqref{eq:_BF_QH} is a special case of Eq.~\eqref{eq:_BF_QH_2} when $\eta^{ab}=\delta^{ab}$. More generally, a Hamiltonian that commutes with Eq.~\eqref{eq:_ang_mom} does not need to have quadratic kinetic energy, as in Eq.~\eqref{eq:_Kohn_Hamiltonian}. As we demonstrate in App.~\ref{app:_QM_Sym}, the coefficient of $f_{0}(\tilde{\theta}(\phi,\theta))$ in Eq.~\eqref{eq:_BF_QH_2} should be replaced by $\sqrt{\det(\mathsf{S}_{2}^{ab})}/\pi$, which satisfies the universal bounds given in Eq.~\eqref{eq:_detS2_bounds}.

Based on our previous discussions, the corner term is expected to diverge as $\theta$ approaches zero. It is straightforward to show that the small-$\theta$ singularity of $f_{0}(\tilde{\theta}(\phi,\theta))$ is given by
\begin{flalign}
f_{0}(\tilde{\theta})\approx\frac{\pi}{\theta}(\sin^{2}(\phi)\eta^{xx}+\cos^{2}(\phi)\eta^{yy}-\sin(2\phi)\eta^{xy}).
\end{flalign}
Therefore, the exact expression  Eq.~\eqref{eq:_BF_QH_2} is consistent with the general result in Eq.~\eqref{eq:_BF_small-angle}. For the orientation-averaged corner term, we have numerically verified that
\begin{flalign}
\int_{0}^{2\pi}\frac{\textrm{d}\phi}{2\pi}f_{0}(\tilde{\theta}(\phi,\theta))=\frac{\Tr(\eta^{ab})}{2}f_{0}(\theta),
\end{flalign}
which confirms the result in Eq.~\eqref{eq:_BF_QH_avg}.

\subsection{Fractional Quantum Hall States}
\label{subsec:_FQHE}





We now present a complementary approach to the localization tensor for quantum Hall states with continuous translational invariance. We will then discuss several conditions for the saturation of the topological bound on $\mathsf{S_2^{ab}}$, with particular emphasis on the role of wavefunction holomorphicity.

As discussed in section \ref{subsec:_bound}, in the presence of continuous translational symmetry, the corner term at small angles is determined purely by the fluctuations of electric polarization $\mathsf{S}_{2}^{ab}$. Moreover, if we also have continuous generalized rotational invariance, it determines the corner term for all angles and orientations. We only focus on $\mathsf{S}_{2}^{ab}$ for the purpose of this section since it plays an important role in the corner term and the investigation of bound on $\mathsf{S}_{2}^{ab}$ is a general and fundamental problem on its own.

$\mathsf{S}_{2}^{ab}$ is a geometric property of the ground state manifold in the twisted boundary conditions parameter space. We follow Ref. \cite{SWM2000}, where a generating function of moments of polarization is introduced, to discuss it without invoking a Hamiltonian unless necessary. One new ingredient in interacting systems is the presence of ground state degeneracy. To account for this, it is useful to generalize Ref. \cite{SWM2000}'s generating function of moments of polarization to an operator in the manifold of ground states. Its matrix elements are given by $C_{nm}(\bm \varPhi, \bm q) = \braket{\Psi_n'|e^{-i q_a \bar r^a N_e}|\Psi_m}$, where $\bar r^a = \frac{1}{N_e} \sum_{j=1}^{N_e} r^a_j$ is the center of mass position that can also be interpreted as dipole moment per particle. Moreover, $\ket{\Psi_n}, \ket{\Psi_n'}$ correspond to the set of ground states with twisted boundary conditions described by the phases $\varPhi_a, \varPhi_a'= \varPhi_a - q_a L$ respectively, where $L$ is the system size. Notice that if $q_a L \ \rm{mod}\ 2\pi = 0$, the boundary conditions are not altered in the $a^{\rm th}$-direction. The static structure factor can be obtained as follows:
\begin{align}
    \mathsf{S}_{2}^{ab} &= -\left.\frac{1}{V N_{\textrm{g}}} \int \frac{\textrm{d}^d\varPhi}{(2\pi)^d}\ \frac{\partial^2 }{\partial q_a \partial q_b}\mathrm{tr}\log C(\bm\varPhi,\bm q) \right|_{\bm q=0}\label{eq:S2_from_C}
\end{align}
where $V=L^d$ is the volume of sample, $N_g$ is the number of ground states and we have averaged over twisted boundary conditions. An advantage of this approach is that $\mathsf{S}_2^{ab}$ is independent of choice of basis for both $\ket{\Psi_n}, \ket{\Psi_m'}$. Further, we show in App. \ref{app:_Quan_Geom} that this expression is identical to the one used in Eq. \eqref{eq:g_deg_gs}.

In App.~\ref{app:_translational_symmetry}, we show how continuous translational symmetry in $d=2$ can be exploited so that $\mathsf{S}_{2}^{ab}$ can be obtained using the following modified generating function that does not modify twisted boundary conditions:
\begin{align}
	\tilde C_{nm}(\bm q) &= \braket{\Psi_n|e^{i\epsilon^{ab}q_a \bar\pi_b/B}|\Psi_m}\nonumber\\
	\mathsf{S}_{2}^{ab} &= -\left.\frac{1}{V N_{\textrm{g}}} \int \frac{\textrm{d}^2\varPhi}{(2\pi)^2}\ \frac{\partial^2 }{\partial q_a \partial q_b}\mathrm{tr}\log \tilde C(\bm q) \right|_{\bm q=0} \label{eq:S2_from_C_tilde}\nonumber\\
    \bar\pi_a &= \sum_{j=1}^{N_e} \pi^j_a
\end{align}
where $\pi_a^j = p_a^j - A_a(\bm r_j)$ is the kinetic momentum of the $j^{\rm th}$ particle.

Now we discuss some explicit examples where the topological bound of Eq. \eqref{eq:_low_bounds}, i.e. $\mathrm{Tr}(\mathsf{S}_2^{ab})/2 \geq \sqrt{\det{\mathsf{S}_2^{ab}}} \geq |\sigma^{xy}|/2$, can be analytically investigated.
These examples involve a unimodular metric $\eta_{ab}$, using which we can define Landau level ladder operators 
\begin{equation}
 \alpha_j=\ell_B \lambda^a \pi_a^j ,  
\end{equation} 
where $\ell_B = \sqrt{1 /|B|}$ is the magnetic length and $\lambda^a$ alongside $\zeta_a$ are complex vectors defined using the metric $\eta_{ab}$ in App.~\ref{app:_Kohn_th}. 

Using the ladder operators, the generating function can be expressed as:
\begin{align}
    \tilde C_{nm}(\bm q) &= e^{-N_e \frac{ \eta^{ab}q_a q_b\ell_B^2}{4}} \nonumber\\
    &\ \ \ \ \ \ \ \times \braket{\Psi_n| e^{-(\bm\lambda.\bm q) \ell_B \sqrt{N_e}\bar\alpha^\dagger}e^{(\bar{\bm\lambda}.\bm q)\ell_B \sqrt{N_e}\bar\alpha} |\Psi_m}\label{eq:C_tilde_alpha}\nonumber\\
    \bar\alpha &= \frac{1}{\sqrt{N_e}} \sum_j \alpha_j.
\end{align}
We emphasize that up to this point $\eta_{ab}$ is an arbitrary metric.

It is evident from Eq. \eqref{eq:C_tilde_alpha} that if $\eta_{ab}$ is such that $\bar\alpha$ annihilates the many-body wavefunctions, we get:
\begin{equation}
     \tilde C_{nm}(\bm q) = e^{-N_e\frac{ \eta^{ab}q_a q_b\ell_B^2}{4}}\delta_{nm},
\end{equation}
which implies
\begin{align}
    \mathsf{S}_2^{ab} &= \frac{|\nu|}{4\pi} \eta^{ab} = \frac{|\sigma^{xy}|}{2} \eta^{ab}.
\end{align}
The equality $\nu=2\pi\sigma^{xy}$ is a result of continuous translational invariance as discussed in App.~\ref{app:_translational_symmetry}.
This saturates the topological bounds of Eqns. \eqref{eq:_detS2_bounds} and \eqref{eq:_mTrS2_bounds}. However, the trace bound of Eq. \eqref{eq:_TrS2_bounds} is saturated only when $\eta_{ab} = \delta_{ab}$.

{As an interesting aside, the condition that $\bar\alpha$ annihilates the ground states determines not only the localization tensor but also the higher moments of polarization, via $\tilde C_{nm}(\bm q)$. These moments appear in the longest wavelength behavior of density operator correlations:
\begin{align}
    \langle \rho(\bm q_1) \cdots \rho(\bm q_m)\rangle_c &\approx (2\pi)^2\delta^{(2)}(\bm q_1 +\cdots + \bm q_m) \nonumber\\
    &\ \ \ \ \ \ \times \mathsf{T}_m^{a_1\cdots a_m} q_{a_1} \cdots q_{a_m}\nonumber\\
    \mathsf{T}_m^{a_1\cdots a_m} &\sim \frac{(-\mathtt{i} N_e)^m}{V} \langle \bar r^{a_1} \cdots  \bar r^{a_m}\rangle_c.
\end{align}
The ``$\sim$'' notation indicates the position operator’s ambiguity on a torus, with moments of $\bar r^a$ derived from the modified generating function.}

To understand the condition for saturation of bound further, let us define a complex coordinate $z_j = \sqrt{2}\zeta_a r^a_j$. The ladder operators in symmetric gauge $A_a(\bm r) = -\epsilon_{ab}B r^b/2$ take the following form:
\begin{align}
    \alpha_j &= \frac{1}{\sqrt{2} \mathtt{i}} \left(2\ell_B \frac{\partial}{\partial z_j^*} + \frac{z_j}{2\ell_B}\right).
\end{align}
When kinetic energy possesses generalized rotational invariance, one can express the single-particle wavefunction as $\psi_j(z_j,z_j^*) = \phi_j(z_j,z_j^*)e^{-\frac{|z_j|^2}{4\ell_B^2}}$. The action of $\alpha_j$ on $\phi_j$ has a simple form:
\begin{align}
    \alpha_j \psi_j(z_j,z_j^*) &= -\mathtt{i}\sqrt{2} \ell_B\left( \frac{\partial}{\partial  z_j^*}\phi_j(z_j, z_j^*) \right)e^{-\frac{|z_j|^2}{4\ell_b^2}}\label{eq:action_of_alpha}.
\end{align}
For a many-body wavefunction, we get:
\begin{align}
    \Psi\left(\{z_j, z_j^*\}\right) &= \Phi\left(\{z_j, z_j^*\}\right) e^{-\sum_{j=1}^{N_e} \frac{|z_j|^2}{4\ell_B^2}}\nonumber\\
    \bar\alpha \Psi &= -\frac{\mathtt{i}\sqrt{2} \ell_B}{\sqrt{N_e}}\left(\sum_{l=1}^{N_e} \frac{\partial \Phi}{\partial z_l^*}\right)e^{-\sum_j \frac{|z_j|^2}{4\ell_B^2}}\nonumber\\
    &= -\frac{\mathtt{i}\sqrt{2} \ell_B}{\sqrt{N_e}} \frac{\partial \Phi}{\partial \bar z^*}\ e^{-\sum_j \frac{|z_j|^2}{4\ell_B^2}}\label{eq:action_of_alpha_bar}.
\end{align}
Thus, a ground state annihilated by the ladder operator $\bar\alpha$, that saturates the bound, possesses holomorphicity or maximal chirality in the center of mass coordinate $\bar z$. In other words, the center of mass coordinate occupies the lowest LL and doesn't mix LLs. 
The holomorphicity has been recently discussed in band systems to give rise to saturation of the topological bounds on quantum geometry~\cite{Claassen2015,Ledwith2020,Mera2021a,Mera2021b,Liu2024}. Here, we focus on many-body states with continuous translational symmetry and many-body quantum metric where holomorphicity in the center of mass coordinate is sufficient.


We now discuss a few examples where we can compute the localization tensor analytically. Except for the last one, all other cases saturate the bound.


\subsubsection{No Landau level mixing\label{subsubsec:upp_LL_unfilled}}
The simplest scenario where the bound is saturated is where only the uppermost Landau level is partially filled in a fermionic model, i.e. no Landau level mixing is present.
To understand the physical situation where this might arise, we assume continuous generalized rotational invariance in the single-particle kinetic energy so that it is purely a function of $\alpha^\dagger_j \alpha_j$ and doesn't mix LLs, i.e. it commutes with angular momentum (see Eq. \eqref{eq:_ang_mom}).
Moreover, the energy of $n^{\rm th}$ Landau level is restricted to be a monotonic function of $n$ and interaction energy scale is assumed to be small compared to the Landau level gap at the Fermi level. One can then approximate the ground state as being made up of completely filled and empty Landau levels for $n<n_0$ and $n>n_0$ respectively. Only the topmost LL labeled by $n=n_0$ might be partially filled.

In this special scenario, the ground state wavefunction is annihilated by $\bar\alpha$ since the LL index of none of the particles can be lowered due to Pauli principle. This would lead to a saturation of the determinant and $\overline{\mathrm{Tr}}$ bounds. We require generalized rotational invariance only in the kinetic energy part of the Hamiltonian, since the interaction induced LL mixing is assumed to be negligible. Note that ``no LL mixing'' is a stronger condition than needed, because one only requires the center of mass position operator to not mix LLs for saturating the bound.

Various model wavefunctions discussed in the literature satisfy this requirement. For example, Laughlin~\cite{Laughlin1983}, Moore-Read~\cite{Moore1991}, Read-Rezayi~\cite{Read1999}, Jain wavefunctions projected to uppermost Landau level~\cite{Jain1989}, and the projected composite-fermion Fermi sea wavefunction~\cite{Rezayi1994} would all saturate the bound. 

An interesting comparison can be made here with the Haldane bound~\cite{Haldane2009}, which provides a lower bound on the fourth order term in projected static structure factor, i.e. $\overline{\mathsf{S}}_4^{abcd}$, by the guiding center Hall viscosity. It was found in Ref. \cite{Kumar2024} that one requires not just projection to a Landau level, but also rotationally invariant model wavefunctions obtained from CFT techniques for saturation of Haldane bound~\cite{Moore1991,Rezayi1994,Hansson2017}. Hence, it requires even stronger notion of maximal chirality (the entanglement spectrum must be completely chiral) than the one discussed in this paper, and involves relative degrees of freedom in addition to the center of mass.

\subsubsection{Invariance under Galilean boosts\label{subsubsec:Galilean_invariance_saturation}}

Beyond the strong condition discussed in previous subsection where Landau level mixing is absent completely, the bound on $\mathsf{S}_2^{ab}$ can also be saturated when the Hamiltonian is invariant under Galilean boosts. The results have already been discussed in \ref{subsec:_Kohn}, here we provide a proof elucidating the holomorphicity of the wavefunction. As shown there, the center of mass degree of freedom decouples from the relative degrees and the corresponding Hamiltonian is given by Eq. \eqref{eq:_Kohn_Hamiltonian_CoM}. In terms of the center of mass ladder operators,
\begin{align}
    H_{\rm CoM} &= \omega_c \left(\bar\alpha^\dagger \bar\alpha + \frac{1}{2}\right)
\end{align}
where $\omega_c = |B|/m$ is the cyclotron frequency. The ground state is thus annihilated by $\bar\alpha$ saturating the bound. This implies the presence of maximal chirality or holomorphicity in the center of mass coordinates. 

The Hamiltonian does not have to possess generalized rotational invariance for this conclusion, in particular the Coulomb metric can be different from the anisotropic mass tensor. Moreover, the particles (can be either fermions or bosons) may occupy multiple Landau levels. Since the center of mass coordinate forms its own LLs, it always occupies the lowest mode in the ground state.


\subsubsection{Unprojected parton wavefunctions}

Surprisingly, unprojected parton wavefunctions~\cite{Jain1989b,Jain1990,Wen1991,Wen1991b,Wen1992,Blok1992} with generalized rotational invariance also have the minimum possible $\sqrt{\det(\mathsf{S}_2^{ab})}$, i.e., $\mathsf{S}_2^{ab} = \frac{|\nu|}{4\pi}\eta^{ab}$, even though they do not necessarily satisfy the criterion in Sec.~\ref{subsubsec:upp_LL_unfilled}. 
We can write an unprojected parton wavefunction as: 
\begin{align} 
\Phi^{\rm pwf}\left(\left\{z_j,z_j^*\right\}\right) &= \prod_{\gamma=1}^{N_{\rm pf}}\Theta_{p_\gamma}\left(\left\{z_j,z_j^*\right\}\right)\label{eq:parton_wavefunction} 
\end{align} 
where $N_{\rm pf}$ is the number of parton factors, $\Theta_p\left(\left\{z_j,z_j^*\right\}\right)$ is the Slater determinant wavefunction of $p$-filled parton LLs (pLLs) without the Gaussian factors. Here $p$ can either be a positive or negative integer. The filling fraction is $\nu^{-1} = \sum_{\gamma=1}^{N_{\rm pf}} p_\gamma^{-1}$~\cite{Jainbook}. Using Eq. \eqref{eq:action_of_alpha_bar}, it is easy to see that the action of $\bar\alpha$ simply distributes over the parton factors so that each $\Theta_{p_\gamma}\left(\left\{z_j,z_j^*\right\}\right)$ can be analyzed independently.

For $p>0$, $\bar\alpha$ annihilates $\Theta_p\left(\left\{z_j,z_j^*\right\}\right)$ since it satisfies the criterion discussed in Sec.~\ref{subsubsec:upp_LL_unfilled}. Essentially, $\alpha_j$ acts as a pLL lowering operator, but the pLL index of none of the partons in $\Theta_p\left(\left\{z_j,z_j^*\right\}\right)$ can be lowered due to the Pauli exclusion principle.

On the other hand, the $p<0$ case is less obvious. To this end, we notice that changing the sign of $p$ is equivalent to changing the sign of the effective magnetic field for the parton and can also be expressed as complex conjugation, i.e., $\Theta_p\left(\left\{z_j,z_j^*\right\}\right) = \left(\Theta_{-p}\left(\left\{z_j,z_j^*\right\}\right)\right)^*$. The LL lowering operator $\alpha_j$ has a different interpretation compared to the $p>0$ case. Here, it corresponds to the intra-pLL angular momentum lowering operator $b_j$: 
\begin{align} 
\alpha_l \Theta_p\left(\left\{z_j,z_j^*\right\}\right) &= \left(\mathtt{i}\sqrt{2}\ell_B\frac{\partial}{\partial z_l}\Theta_{-p}\left(\left\{z_j,z_j^*\right\}\right)\right)^*\nonumber\\ 
&= \left(b_l \Theta_{-p}\left(\left\{z_j,z_j^*\right\}\right)\right)^*.
\end{align} 
Since the parton factor $\Theta_{-p}\left(\left\{z_j,z_j^*\right\}\right)$ at $p<0$ has the minimum possible total angular momentum in each pLL in the absence of quasi-hole excitations, 
we conclude that $\bar\alpha$ annihilates it as well.
As a result, the unprojected parton wavefunction of Eq. \eqref{eq:parton_wavefunction} is entirely annihilated by the $\bar\alpha$ operator, saturating the topological lower bound for $\sqrt{\det(\mathsf{S}_2^{ab})}$.

This class of wavefunctions provides several useful models for fractional quantum Hall states. A set of prominent ones are unprojected Jain states~\cite{Jain1989} at $\nu= \frac{m}{mr\pm 1}$. These correspond to ``$r$'' number of parton factors with $p=1$ and one extra factor with $p=\pm m$. Other examples include non-abelian FQH states~\cite{Wen1991,Wen1998,Wen1999,Balram2018,Balram2019}.
It would be interesting to confirm these expectations using numerical techniques on geometries other than the plane, especially the non-chiral cases that involve negative effective magnetic fields for at least some of the partons.

\subsubsection{Composite-fermion Fermi sea wavefunctions}
\label{subsubsec:_CFFS_saturation}

We can also analyze the unprojected composite-fermion Fermi sea (CFFS) wavefunctions~\cite{Rezayi1994} that possess generalized rotational invariance using the methods of this section. At $\nu=1/m$, they are described by:
\begin{align}
    \Phi^{\rm CFFS}_{\frac{1}{m}}\left(\left\{z_j,z_j^*\right\}\right) &= \left(\prod_{i<j}(z_i-z_j)^{m}\right) \det \left(e^{\mathtt{i}\frac{k^{j*} z_l + k^j z_l^*}{2}}\right).
\end{align}
The second factor is the Slater determinant that describes a filled Fermi sea of non-interacting CFs and $k^j = k_x^j + \mathtt{i}k_y^j$ are the corresponding complex wavevectors. It is straightforward to see that:
\begin{align}
    \bar\alpha \Phi^{\rm CFFS}_{\frac{1}{m}} &= \frac{\ell_B}{\sqrt{2 N_e}}\left(\sum_{j=1}^{N_e}k^j\right) \Phi^{\rm CFFS}_{\frac{1}{m}}\nonumber\\
    &= 0.
\end{align}
The last line is obtained using the fact that in the presence of generalized rotational invariance, the sum of all wavevectors is zero. Thus, CFFS wavefunctions also saturate the topological bound for $\sqrt{\det(\mathsf{S}_2^{ab})}$. Our analytical derivation explains the recent numerical results of Ref. \cite{CFLEE3} where both the projected and unprojected isotropic CFFS wavefunctions were found to be consistent with Eq. \eqref{eq:_BF_QH}.

\subsubsection{Decoupled Landau levels}

We now discuss one explicit case where the bound is not saturated. Consider a wavefunction of fermions where multiple Landau levels are (partially) occupied and a QH state is formed within each LL. Different LLs are assumed to be decoupled, i.e. the many-body wavefunction is a direct product of the QH states of all LLs before anti-symmetrization. We further require generalized rotational invariance so that there is no LL mixing at the single particle level. The $n^{\rm th}$ single-particle LL wavefunction for $j^{\rm th}$ particle $\ket{j;n,k}$ satisfies $\alpha_j^\dagger\alpha_j \ket{j;n,k} = n \ket{j;n,k}$. Here, $k$ is taken to be the the eigenvalue of magnetic momentum along $x$-direction on the torus. Using second quantization representation, we can write:
\begin{align}
    \bar\alpha &= \frac{1}{\sqrt{N_e}}\sum_{n=1}^{\infty}\sum_k \sqrt{n}\ c^\dagger_{n-1,k} c_{n,k} 
\end{align}
where $c_{n,k}$ annihilates a particle in the state $\ket{n,k}$.

For the decoupled LLs case, particle number and momentum of each LL are independently conserved. The following expectation value then takes a simple form:
\begin{align}
    \langle \bar\alpha^\dagger \bar\alpha\rangle &= \frac{1}{N_e}\sum_{n,n',k,k'} \sqrt{nn'}\left\langle c^\dagger_{n,k} c_{n-1,k} c^\dagger_{n'-1,k'} c_{n',k'}\right\rangle\nonumber\\
    &= \frac{1}{N_e}\sum_{n,k} n \left\langle\hat n_{n,k}\right\rangle \left\langle1-\hat n_{n-1,k}\right\rangle
    \nonumber\\
    &= \frac{1}{|\nu|} \sum_{n=1}^\infty n|\nu_n| (1-|\nu_{n-1}|)
\end{align}
where $\nu_n$ is the filling fraction of $n^{\rm th}$ LL such that $\sum_n \nu_n = \nu$. The correlation function in second line has only a disconnected contribution.

Using Eqns. \eqref{eq:C_tilde_alpha}, \eqref{eq:S2_from_C_tilde}, we find that in the presence of generalized rotational invariance:
\begin{align}
    \mathsf{S}_2^{ab} &= \left(\langle \bar\alpha^\dagger \bar\alpha\rangle + \frac{1}{2}\right)\frac{|\nu|}{2\pi} \eta^{ab} \nonumber\\
    &=\left(\frac{|\nu|}{2} + \sum_{n=1}^{\infty} n|\nu_n|(1-|\nu_{n-1}|)\right) \frac{\eta^{ab}}{2\pi}
\end{align}
If the second term is not zero, the topological lower bounds on $\mathsf{S}_2^{ab}$ are not satisfied with equality. A simple non-interacting realization of this is the $\nu=1$ state with first Landau level filled and zeroth empty. It could arise in a band-inverted system where the kinetic energy has a Mexican-hat shape.



This class of examples demonstrates that the bound saturation is achieved under special conditions. Since they can be thought of as extreme cases of LL mixing, we expect the same for more realistic situations where interaction energy is not small compared to the Landau level gap in either integer or fractional QH states and LL mixing is present (the Galilean boost invariant case discussed in subsection \ref{subsubsec:Galilean_invariance_saturation} is an exception).

An important implication in the context of corner term of bipartite fluctuations is that even when the system possesses continuous translational and generalized rotational symmetries, the coefficient of the corner angle functions in Eqns. \eqref{eq:_BF_QH}, \eqref{eq:_BF_QH_2} is not $|\sigma^{xy}|/2\pi$, but is bounded below by it.

At present, we are not aware of other cases that saturate the bound on $\mathsf{S}_{2}^{ab}$. It is however clear that our examples are fine tuned such that the ground state is annihilated by a suitably defined Landau level lowering operator and thus possesses holomorphicity in the center of mass coordinate. 
It would be especially interesting to find examples where neither Galilean boost nor generalized rotational invariance is present.

As already discussed in section \ref{sec:_Chern_insulators}, non-interacting Chern insulators, that lack continuous translational invariance, do not saturate the bound generally. Similarly, we do not expect generic fractional QH states to do so in such settings either. Nevertheless, at the non-interacting level, ideal bands recently studied in literature are known to saturate the topological bounds~\cite{Wang2021,Liu2024}. Exploring FQH states, both projected and not projected to these bands, is an interesting problem in this context when Berry curvature is distributed non-uniformly over the Brillouin zone.

\section{Composite Fermi Liquids}
\label{sec:_CFL}


In this section, we turn our attention to the compressible QH state at half-filling. The physical properties of this gapless state can be understood by considering the Fermi surface state of composite fermions~\cite{HLR1993,son2015}, which form by attaching two flux quanta to each electron. Recently, similar physics has been proposed for the half-filled Chern band in TMD Mori\'{e} materials without a magnetic field~~\cite{tmdcfl1,tmdcfl2}. Bipartite fluctuations in such non-Fermi liquid states have been studied in Refs.~\cite{BF_CFS,disop_FS,CFLEE3}.

Conventional electron metals, as described by Landau Fermi liquid theory, are characterized by finite compressibility, finite Drude weight in the clean limit, and a divergent localization tensor~\cite{Kohn1964, SWM2000,Resta2002Rev, Resta2011Rev}. Composite Fermi liquids (CFLs), while also compressible, exhibit dramatic differences compared to conventional Fermi liquids. The vanishing of the Drude weight can be easily understood using the Ioffe-Larkin decomposition rule
\begin{flalign}
\sigma_{e}^{-1}=\sigma_{\textrm{cf}}^{-1}+\sigma_{\textrm{cs}}^{-1},
\label{eq:_I-L_CFL}
\end{flalign}
where $\sigma_{e}^{ab}$ is the electron conductivity tensor, $\sigma_{\textrm{cf}}^{ab}$ is the conductivity of composite fermions, and $\sigma_{\textrm{cs}}^{ab}=(\nu/2\pi)\varepsilon^{ab}$ comes from the Chern-Simons term~\cite{HLR1993}. Although the mean-field state of composite fermions may exhibit a singularity at low frequency, i.e., $\sigma_{\textrm{cf}}^{xx}\supset D_{\textrm{cf}}(\delta(\omega)+\frac{\mathtt{i}}{\pi\omega})$, the longitudinal conductivity of electrons behaves as $\sigma_{e}^{xx}\sim\mathtt{i}\omega$, which vanishes in the limit $\omega\rightarrow0$. This property has recently been highlighted in the context of the half-filled Chern band in Ref.~\cite{cfl_drude}.

Another remarkable feature of CFLs is their finite localization tensor, i.e., a finite many-body quantum metric. This is linked to a finite corner term in bipartite charge fluctuations. As first pointed out in Ref.~\cite{BF_CFS}, the formula in Eq.~\eqref{eq:_BF_QH} holds for the original Halperin-Lee-Read (HLR) theory~\cite{HLR1993}. Recently, this has been numerically verified by Monte Carlo simulations in Ref.~\cite{CFLEE3} using the Rezayi-Read wavefunction, considering CFLs for fermions at $\nu=1/2$ and $\nu=1/4$, as well as CFLs for bosons at $\nu=1$ and $\nu=1/3$ fillings. Here, we further clarify the results of previous studies~\cite{BF_CFS,CFLEE3}. The angle function in Eq.~\eqref{eq:_angle_function_0} is guaranteed by continuous rotational invariance, as higher-order terms in Eq.~\eqref{eq:_BF_corner_expd} vanish automatically. In Ref.~\cite{CFLEE3}, the bound on the coefficient is saturated due to the fulfillment of the holomorphicity condition for the CF wavefunctions as described in Sec.~\ref{subsubsec:_CFFS_saturation}. However, unsaturated cases are expected to occur in the absence of Galilean invariance when Landau-level mixing is also present. In Ref.~\cite{BF_CFS}, the bound saturation is verified using the random phase approximation (RPA) based on the HLR theory~\cite{HLR1993} for isotropic systems. Nevertheless, as we will examine in detail in Sec.~\ref{subsec:_Ani_CFL}, the standard RPA is generally insufficient for anisotropic systems.


\subsection{Anisotropic CFLs}
\label{subsec:_Ani_CFL}

We extend the discussions from Ref.~\cite{BF_CFS} to anisotropic systems with continuous translation. We begin by considering the half-filled Landau level with the generalized rotational symmetry described in Sec.~\ref{subsec:_QM_Sym}. Such a system can be realized by the UV Hamiltonian given in Eq.~\eqref{eq:_Kohn_Hamiltonian}, where the Galilean metric $\eta_{ab}$ and the Coulomb metric $\widetilde{\eta}_{ab}$ are identified. Let us consider an anisotropic version of the HLR theory~\cite{HLR1993} for $\nu=1/2$
\begin{widetext}
\begin{flalign}
\mathcal{L}=\psi^{\dagger}\left(\textrm{D}_{\tau}-\frac{\widehat{\eta}^{ab}}{2m_{*}}\textrm{D}_{a}\textrm{D}_{b}-\mu_{*}\right)\psi-\frac{1}{2}\frac{\mathtt{i}\varepsilon^{\mu\nu\rho}}{4\pi}(a_{\mu}-A_{\mu})\partial_{\nu}(a_{\rho}-A_{\rho})+(\textrm{interactions}),
\label{eq:_HLR_action}
\end{flalign}
\end{widetext}
where $\textrm{D}_{\mu}=\partial_{\mu}-\mathtt{i}a_{\mu}$ is the gauge covariant derivative. Another unimodular metric $\widehat{\eta}_{ab}$ is introduced to describe the mass anisotropy of the composite fermions $\psi$, and its inverse is denoted by $\widehat{\eta}^{ab}$. The effective mass $m_{*}$ generally differs from the bare mass $m$ of the electrons, and the chemical potential $\mu_{*}$ ensures that the density of composite fermions matches that of the electrons. It is convenient to introduce a spacetime unimodular metric 
\begin{flalign}
\widehat{\eta}_{\mu\nu}=\left(\begin{array}{ccc}
1 & 0 & 0\\
0 & \widehat{\eta}_{xx} & \widehat{\eta}_{xy}\\
0 & \widehat{\eta}_{xy} & \widehat{\eta}_{yy}
\end{array}\right),
\end{flalign}
and consider the coordinate transformations
\begin{flalign}
\tilde{q}_{I}&=\widehat{e}_{I}^{\mu}q_{\mu},\qquad\tilde{r}^{I}=\widehat{e}_{\mu}^{I}r^{\mu},\nonumber\\\tilde{a}_{I}&=\widehat{e}_{I}^{\mu}a_{\mu},\qquad\tilde{A}_{I}=\widehat{e}_{I}^{\mu}A_{\mu}.
\label{eq:_coord_tran_IR}
\end{flalign}
Here, the vielbeins $\widehat{e}_{I}^{\mu}$ and $\widehat{e}_{\mu}^{I}$ are introduced to diagonalize $\widehat{\eta}^{\mu\nu}=\widehat{e}_{I}^{\mu}\widehat{e}_{J}^{\nu}\delta^{IJ}$ and its inverse $\widehat{\eta}_{\mu\nu}=\widehat{e}_{\mu}^{I}\widehat{e}_{\nu}^{J}\delta_{IJ}$. Following the coordinate transformation in Eq.~\eqref{eq:_coord_tran_UV}, the isotropic UV system described by Eq.~\eqref{eq:_Kohn_Hamiltonian} should correspond to an isotropic CFL theory. This is presumably related to the HLR theory Eq.~\eqref{eq:_HLR_action} after the coordinate transformation in Eq.~\eqref{eq:_coord_tran_IR}. As we will demonstrate self-consistently below, setting the UV and IR unimodular metrics equal $\eta_{ab}=\widehat{\eta}_{ab}$ satisfies the result in Eq.~\eqref{eq:_S2_Kohn}, as required by the generalized Kohn theorem.

In the new coordinate system, the first term in Eq.~\eqref{eq:_HLR_action} becomes isotropic, while the Chern-Simons term remains invariant. The standard RPA approach~\cite{HLR1993} results in 
\begin{flalign}
\tilde{S}(\tilde{\boldsymbol{q}})=\int\frac{\textrm{d}\tilde{q}_{\tau}}{2\pi}\frac{1}{\Pi_{\psi}^{\tau\tau}(\tilde{q})^{-1}-U(\tilde{\boldsymbol{q}})-\frac{(4\pi)^{2}}{|\tilde{\boldsymbol{q}}|^{2}}\Pi_{\psi}^{TT}(\tilde{q})},
\label{eq:_HLR_RPA}
\end{flalign}
where the longitudinal and transverse components of the response function for composite fermions are 
\begin{flalign}
\Pi_{\psi}^{\tau\tau}(\tilde{q})&=-\mathscr{D}_{F}\left(1-\frac{|\tilde{q}_{\tau}|}{\sqrt{\tilde{q}_{\tau}^{2}+(\tilde{v}_{F}\tilde{\boldsymbol{q}})^{2}}}\right),\nonumber\\\Pi_{\psi}^{TT}(\tilde{q})&=\mathscr{D}_{F}\frac{|\tilde{q}_{\tau}|\sqrt{\tilde{q}_{\tau}^{2}+(\tilde{v}_{F}\tilde{\boldsymbol{q}})^{2}}-\tilde{q}_{\tau}^{2}}{|\tilde{\boldsymbol{q}}|^{2}}.
\end{flalign}
Here, $\tilde{v}_{F}$ is the fermi velocity, and $\mathscr{D}_{F}=\tilde{k}_{F}/(2\pi\tilde{v}_{F})$ represents the density of states at the fermi level. The value of the fermi momentum $\tilde{k}_{F}$ is determined by the Luttinger theorem $\tilde{V}_{\textrm{FS}}/(2\pi)=\nu(2\pi\ell_{B}^{2})$, where the fermi-surface volume is $\tilde{V}_{\textrm{FS}}=\pi\tilde{k}_{F}^{2}$. Considering the Coulomb potential $U(\tilde{\boldsymbol{q}})\sim1/|\tilde{\boldsymbol{q}}|$, the dipolar interaction $U(\tilde{\boldsymbol{q}})\sim|\tilde{\boldsymbol{q}}|$, and the Gaussian potential $U(\tilde{\boldsymbol{q}})\sim e^{-s^{2}|\tilde{\boldsymbol{q}}|^{2}/2}$, we find the static structure factor
\begin{flalign}
\tilde{S}(\tilde{\boldsymbol{q}})=\frac{|\nu|}{4\pi}|\tilde{\boldsymbol{q}}|^{2}+c_{U}|\tilde{\boldsymbol{q}}|^{3}\log(1/|\tilde{\boldsymbol{q}}|)+\mathscr{O}(|\tilde{\boldsymbol{q}}|^{4}),
\label{eq:_SSF_CFL}
\end{flalign}
where the coefficient $c_{U}$ is given by 
\begin{flalign}
c_{U}=\begin{cases}
1/(8\pi^{2}\tilde{k}_{F}), & \textrm{Coulomb }U(\tilde{\boldsymbol{q}})\sim1/|\tilde{\boldsymbol{q}}|\\
1/(4\pi^{2}\tilde{k}_{F}), & \textrm{Dipolar }U(\tilde{\boldsymbol{q}})\sim|\tilde{\boldsymbol{q}}|\\
1/(4\pi^{2}\tilde{k}_{F}), & \textrm{Gaussian }U(\tilde{\boldsymbol{q}})\sim e^{-\frac{s^{2}|\tilde{\boldsymbol{q}}|^{2}}{2}}
\end{cases}
\label{eq:_HLR_S3}
\end{flalign}
Since  $A_{\tau}=\tilde{A}_{\tau}$ in Eq.~\eqref{eq:_coord_tran_IR}, the density response in the original coordinate system gives $S(\boldsymbol{q})=(\nu/4\pi)\widehat{\eta}^{ab}q_{a}q_{b}$. Identifying $\widehat{\eta}^{ab}$ with $\eta^{ab}$ ensures that Eq.~\eqref{eq:_S2_Kohn} is satisfied. Additionally, as discussed in Ref.~\cite{BF_CFS} the contribution $|\tilde{\boldsymbol{q}}|^{3}\log(1/|\tilde{\boldsymbol{q}}|)$ from gapless modes affects only the leading-order boundary-law term and does not influence the corner term in Eq.~\eqref{eq:_BF_QH_2}.



More generally, when the Galilean metric $\eta_{ab}$ and the Coulomb metric $\widetilde{\eta}_{ab}$ differ in the Hamiltonian Eq.~\eqref{eq:_Kohn_Hamiltonian}, the interaction potential $U(\tilde{\boldsymbol{q}})$ in Eq.~\eqref{eq:_HLR_RPA} becomes anisotropic and is a function of $\widetilde{\eta}^{ab}\widehat{e}_{a}^{I}\widehat{e}_{b}^{J}\tilde{q}_{I}\tilde{q}_{J}$. Revisiting the RPA calculation in Eq.~\eqref{eq:_HLR_RPA}, one finds the same leading term $(\nu/4\pi)|\tilde{\boldsymbol{q}}|^{2}$ as in Eq.~\eqref{eq:_SSF_CFL}, which is independent of the Coulomb metric $\widetilde{\eta}_{ab}$. From the constraint in Eq.~\eqref{eq:_S2_Kohn} imposed by the Kohn theorem, one may again conclude that $\widehat{\eta}_{ab}=\eta_{ab}$, indicating that the composite fermion Fermi surface is identical to the anisotropic electron Fermi surface. This result is similar to the earlier findings in Ref.~\cite{AniCFL1}, which were based on the asymptotic Ward identities of Chern-Simons theory. However, it contrasts with later studies that carefully incorporate the Landau-level projection. Considering the Hamiltonian Eq.~\eqref{eq:_Kohn_Hamiltonian} with $\eta^{ab}=\textrm{diag}(v_{e},1/v_{e})$ and various forms of isotropic interactions $U(r)$ (i.e., $\widetilde{\eta}_{ab}=\delta_{ab}$), it was found that $\widehat{\eta}^{ab}=\textrm{diag}(v_{\textrm{cf}},1/v_{\textrm{cf}})$ satisfies~\cite{AniCFL2,AniCFL3,AniCFL4,AniCFLrev}
\begin{flalign}
v_{\textrm{cf}}=\begin{cases}
v_{e}^{0.493}, & \textrm{Coulomb }U(r)\sim1/r\\
v_{e}^{0.795}, & \textrm{Dipolar }U(r)\sim1/r^{3}\\
\left(\frac{\ell_{B}^{2}v_{e}+s^{2}}{\ell_{B}^{2}/v_{e}+s^{2}}\right)^{\frac{1}{2}}, & \textrm{Gaussian }U(r)\sim e^{-\frac{r^{2}}{2s^{2}}}
\end{cases},
\end{flalign}
where the Gaussian potential result can be derived analytically~\cite{AniCFL2}, and the results for Coulomb and dipolar interactions are determined by density matrix renormalization group (DMRG) simulations~\cite{AniCFL4,AniCFLrev}. Assuming the validity of the RPA calculation in Eq.~\eqref{eq:_HLR_RPA}, only the terms starting from the second in Eq.~\eqref{eq:_SSF_CFL} are affected by the different forms of $U(r)$ as described in Eq.~\eqref{eq:_HLR_S3}. We aim to explore in future studies how to modify Eq.~\eqref{eq:_HLR_action} to account for the fact that $\widehat{\eta}^{ab}$ is generally less anisotropic than $\eta^{ab}$. One may need to consider the anisotropic generalization of the modified RPA~\cite{HLR_MRPA}, which takes into account the Landau parameter.



\subsection{Periodic Potentials}
\label{subsec:_Latt_CFL}

\begin{table}
\begin{tabular}{|c|c|c|}
\hline 
 & FL & CFL\tabularnewline
\hline 
Drude weight & finite & zero\tabularnewline
\hline 
Localization tensor & divergent & finite\tabularnewline
\hline 
Corner charge fluctuation & N/A & finite\tabularnewline
\hline 
\end{tabular}

\caption{The distinct behaviors of the Drude weight~\cite{cfl_drude}, the localization tensor, and the corner charge fluctuations~\cite{BF_CFS}
across the bandwidth-tuned transition between a Landau Fermi liquid (FL)
and a composite Fermi liquid (CFL).}
\label{tab:_cflfl}

\end{table}

Another type of interesting model involves adding a periodic background potential to the (isotropic) Hamiltonian in Eq.~\eqref{eq:_Kohn_Hamiltonian}. This serves as a simplified model for the physics of periodically modulated Landau levels, inspired by TMD Moir\'{e} materials~\cite{tmdll1,tmdll2}. The periodic potential typically breaks the continuous rotational symmetry down to a discrete one, and the formalism for corner charge fluctuations developed in Sec.~\ref{subsec:_Fourier} and Sec.~\ref{subsec:_Latt} is applicable in this context. Unlike flat Landau levels, the energy levels in this model are dispersive, and the bandwidth can be tuned by adjusting the periodic potential (i.e., changing the displacement field in the experiment). It is theoretically understood that this will drive a transition between fermi liquids and composite fermi liquids~\cite{cflfl1,cflfl2,cflfl3,cfl_drude,BF_CFS}. The distinct behaviors of the Drude weight across this transition have recently been discussed in Ref.~\cite{cfl_drude}. Here, we emphasize the different behaviors of bipartite charge fluctuations~\cite{disop_FS,BF_CFS} in both the leading term as well as the corner term, which are related to transport properties through the value of the localization tensor $\mathsf{S}_{2}^{ab}$, determined by the fluctuation-dissipation theorem-type sum rule in Eq.~\eqref{eq:_SWM_sum}. As explained in Ref.~\cite{uni_corner,disop_FS,BF_CFS}, the charge correlation in the FL phase decays too slowly in space, resulting in a leading term $\sim L_{\textrm{A}}\log(L_{\textrm{A}})$, along with a subleading term $\sim L_{\textrm{A}}$ that is sensitive to the entire geometry of the real-space subregion (in FIG.~\ref{fig:_corner}) and the shape of the fermi surface. Consequently, a local corner contribution like Eq.~\eqref{eq:_BF_corner} is no longer well-defined. This non-local feature is linked to the power-law divergence of the localization tensor $\mathsf{S}_{2}^{ab}\sim1/q$ at long wavelengths. The different behaviors across the transition are summarized in TABLE.~\ref{tab:_cflfl}. 

An intriguing possibility is that this transition is actually continuous~\cite{cflfl1,cflfl2,cflfl3}. We predict that the localization tensor exhibits the universal behavior
\begin{flalign}
\frac{\textrm{Tr}(\mathsf{S}_{2}^{ab})}{2}=\frac{\pi}{2}C_{J}(-\upsilon_{*} )\log|w-w_{c}|,
\label{eq:_S2_cflfl}
\end{flalign}
when approaching the critical point from the CFL phase, where $w$ denotes the bandwidth, and $w_{c}$ is its critical value. The universal numbers $\upsilon_{*} $ and $C_{J}$ are identified as the correlation length exponent and the current central charge associated with the transition~\cite{cflfl1} between the superfluid and the bosonic Laughlin state at $\nu=1/2$. Their values, determined under the large-$N$ expansion, are available in Refs.~\cite{wufisher,cflfl1}. Despite the lack of conformal symmetry in the entire system, the bipartite fluctuations at the critical point are expected to follow Eq.~\eqref{eq:_BF_CFT}, as a consequence of Eq.~\eqref{eq:_S2_cflfl} (see also Ref.~\cite{BF_CFS}).

\section{Summary and Discussion}
\label{sec:_Summary}

In this paper, we present a comprehensive study of corner charge fluctuations in general anisotropic many-body systems. The special cases of isotropic systems discussed in Refs.~\cite{Dirac_log, wulog,uni_corner,chengdisop,BF_CFS} are understood within a unified framework from the perspective of many-body quantum geometry.
We demonstrate that, in general, the corner term depends on the corner angle, the absolute orientation of the subregion, and other microscopic details of the system. For continuous (long-wavelength) models, we provide a harmonic expansion representation, given by Eq.~\eqref{eq:_BF_corner_expd}, which consists of generalized universal angle functions Eq.~\eqref{eq:_angle_functions} and their non-universal coefficients Eq.~\eqref{eq:_angle_functions_coef}. In the limit of small opening angle, we find that the corner term is fully determined by the quantum metric (see Eq. \eqref{eq:_BF_small-angle}).
Various properties of the corner term, as understood from the harmonic expansion in Sec.~\ref{sec:_Dis_Sym}, are further confirmed by numerical studies of lattice models in Sec.~\ref{sec:_Chern_insulators} and by the analytical results for an exactly solvable case in Sec.~\ref{subsec:_QM_Sym} (also see App.~\ref{app:_QM_Sym}).


We further clarify the saturation of the universal bounds for the quantum metric, a fundamental problem in its own right, which is also related to the small-angle limit in Eq.~\eqref{eq:_BF_small-angle} and the orientation-averaged expression in Eq.~\eqref{eq:_BF_avg}. Our understanding is based on an anisotropic generalized version of the Kohn theorem and the holomorphic properties of many-body wavefunctions (see Sec.~\ref{sec:_Ani_LL}, App.~\ref{app:_Kohn_th}, and App.~\ref{app:_translational_symmetry} for more details).

For simplicity, we focused on examples with electric charge conservation in Sec.~\ref{sec:_Chern_insulators}, Sec.~\ref{sec:_Ani_LL}, and Sec.~\ref{sec:_CFL}. However, the general discussions concerning bipartite fluctuations discussed in Sec.~\ref{sec:_Corner_Rev}, Sec.~\ref{sec:_Dis_Sym}, and App.~\ref{app:_QM_Sym} rely solely on the existence of a global U(1) symmetry. These results are directly applicable to systems where the total spin in one direction is conserved, such as quantum spin Hall insulators and chiral topological superconductors, where the quantum metric is defined through the insertion of spin-U(1) flux (see App.~\ref{app:_Quan_Geom}). Additionally, our findings in Sec.~\ref{sec:_Ani_LL} could provide insight into the conditions for bound saturation of the spin quantum metric in certain correlated systems analogous to Landau levels, such as chiral spin liquids~\cite{csl2,csl3}. Furthermore, the universal behavior described by Eq.~\eqref{eq:_S2_cflfl} finds its analog in the spin structure factor at continuous Mott transitions~\cite{BF_CFS}.


In the context of free fermions, it is well-known that entanglement entropies are entirely determined by charge fluctuations $\mathcal{N}_{\textrm{A}}^{[2]}$ and higher-order cumulants~\cite{EEBF1,EEBF2,EEBF3,EEBF4,EEBF6,EEBF7,EEBF9}. Given that the corner term in $\mathcal{N}_{\rm A}^{[2]}$ generally exhibits non-universal dependence on the opening angle and the orientation, it is plausible that similar results hold for the entanglement entropy. See Ref.~\cite{corner_QM} for a numerical analysis of the corner contribution to entanglement entropies in band insulators. Additionally, see Ref.~\cite{EE_QM} for recent related discussions that do not involve corners.

The corner contributions to the U(1) disorder operator (as defined in Eq.~\eqref{eq:_dis_op}), also known as full-counting statistics, have been investigated for integer QH and Laughlin states in Ref.~\cite{FCS_corner}. The higher-order cumulants $\mathcal{N}_{\textrm{A}}^{[m]}$, as defined in Eq.~\eqref{eq:_cumulant}, have been found to exhibit distinct common features when $m$ is even and $m$ is odd. According to our generating function approach discussed in Sec.~\ref{subsec:_FQHE}, wavefunction holomorphicity imposes strong constraints on multi-point density correlations in the long-wavelength limit. We hope that future studies will further elucidate the role of spacetime symmetries and wavefunction properties in higher-order cumulants, particularly in distinguishing between universal and non-universal structures, as we have done for bipartite fluctuations.

{The boundary-bulk correspondence has proven valuable in understanding various aspects of gapped phases in $2+1$ dimensions. Notably, the entanglement spectrum in many topological phases corresponds to the energy spectrum of the edge CFT~\cite{Li-Haldane_2008}. This relationship has led to the prediction of a universal subleading term for discrete-symmetry disorder operators without corners~\cite{TDP_2022}. However, it remains unclear whether our results can be fully explained within the framework of edge theory. Despite the efforts made in Ref.~\cite{MultipartiteChiral2024}, accurately reproducing the universal angle function in the edge CFT approach remains an open question.}

Exploring fluctuations of ordinary and generalized symmetry charges in three spatial dimensions presents intriguing possibilities. On one hand, there are various singular geometries to examine, including corners, cones, and trihedral vertices. Some interesting angle dependencies for non-interacting Dirac systems have already been discussed in Ref.~\cite{Dirac_log}. On the other hand, nontrivial gapped phases offer even richer possibilities, such as fractons and 3D topological orders. It would be interesting to understand the universal and non-universal features of shape dependence and to investigate whether there are constraints from topological properties.




\section*{Acknowledgment}

We thank Luca Delacrétaz, Lukasz Fidkowski, Michael Levin, Hridis Pal, Karlo Penc, Dam Thanh Son, Pok Man Tam, Jie Wang, and Cenke Xu for related discussions. X.W. and P.K. were supported in part by the Leo P. Kadanoff Fellowship at the University of Chicago, the Simons Collaboration on Ultra-Quantum Matter, which is a grant from the Simons Foundation (651440), and the Simons Investigator award (990660). M.C. was partially supported by NSF grant DMR-1846109.

{\it Note added:} While preparing the manuscript, two papers~\cite{corner_QM,Fu2024-3} on related topics appeared whose results overlap with ours. In particular, the small-angle limit in Eq.~\eqref{eq:_BF_small-angle_Latt} for the corner charge fluctuations in lattice models was also found in Ref.~\cite{corner_QM}, and the bounds in Eq.~\eqref{eq:_TrS2_bounds} and Eq.~\eqref{eq:_detS2_bounds} for the quantum weight (i.e., the many-body quantum metric) were also obtained in Ref.~\cite{Fu2024-3}. The updated version of Ref.~\cite{corner_QM}, which addresses the small-angle limit of interacting systems, will appear in the same arXiv listing.

\begin{widetext}

\appendix

\section{Harmonic Expansion}
\label{app:_Fourier}

In this appendix, we provide additional technical details regarding the results discussed in Sec.~\ref{subsec:_Fourier}. The static structure factor Eq.~\eqref{eq:_SSF_1} has the harmonic expansion
\begin{flalign}
S_{+}(\boldsymbol{q})=\sum_{n\in\mathbb{Z}}S_{n}(q)e^{\mathtt{i}nN\varphi},
\label{eq:_SSF_3}
\end{flalign}
where $\boldsymbol{q}=q(\cos\varphi,\sin\varphi)$, and the complex-valued coefficients are
\begin{flalign}
S_{n}(q)=\frac{1}{2\pi}\int_{0}^{2\pi}\textrm{d}\varphi\, S_{+}(\boldsymbol{q})e^{-\mathtt{i}nN\varphi}.
\label{eq:_SSF_3_coef}
\end{flalign}
Under Fourier transformations, $S_{n}(r)$ in Eq.~\eqref{eq:_SSF_2} are related to $S_{n}(q)$ as follows
\begin{flalign}
S_{n}(q)&=2\pi\mathtt{i}^{nN}\int_{0}^{+\infty}\textrm{d}r\,rS_{n}(r)J_{nN}(qr),\nonumber\\S_{n}(r)&=\frac{1}{2\pi\mathtt{i}^{nN}}\int_{0}^{+\infty}\textrm{d}q\,qS_{n}(q)J_{nN}(qr),
\label{eq:_radial_Fourier}
\end{flalign}
where $J_{m}(z)$ are the Bessel functions. 

\subsection{Universal Angle Functions}

Considering the subsystem shown in FIG.~\ref{fig:_corner}, we utilize the method described in Ref.~\cite{uni_corner} to isolate the corner term. We start by noting that the bipartite fluctuations in systems with global charge conservation can be rewritten as
\begin{flalign}
\mathcal{N}_{\textrm{A}}^{[2]}=-\int_{\textrm{A}}\textrm{d}^{2}\boldsymbol{r}_{1}\int_{\bar{\textrm{A}}}\textrm{d}^{2}\boldsymbol{r}_{2}\,S_{+}(\boldsymbol{r}_{1}-\boldsymbol{r}_{2}),
\end{flalign}
where $\bar{\textrm{A}}$ denotes the complement of region $\textrm{A}$. The corner term can then be extracted using the expression
\begin{flalign}
\gamma(\phi,\theta)=-\frac{1}{2}(\mathcal{N}_{\textrm{A}}^{[2]}+\mathcal{N}_{\textrm{C}}^{[2]}-\mathcal{N}_{\textrm{AB}}^{[2]}-\mathcal{N}_{\textrm{AD}}^{[2]}),
\end{flalign}
which leads to Eq.~\eqref{eq:_BF_corner}. According to Eq.~\eqref{eq:_SSF_2}, it has the following series expansion 
\begin{flalign}
\gamma(\phi,\theta)=-\sum_{n\in\mathbb{Z}}\int\textrm{d}^{2}\boldsymbol{r}\,\mathbbm{n}(\boldsymbol{r})S_{n}(r)e^{\mathtt{i}nN\varphi},
\end{flalign}
where $\boldsymbol{r}=r(\cos\varphi,\sin\varphi)$, and the function $\mathbbm{n}(\boldsymbol{r})$ is given by 
\begin{flalign}
\mathbbm{n}(\boldsymbol{r})=\int_{\textrm{B}}\textrm{d}^{2}\boldsymbol{r}_{1}\int_{\textrm{D}}\textrm{d}^{2}\boldsymbol{r}_{2}\,\delta^{2}(\boldsymbol{r}_{1}-\boldsymbol{r}_{2}-\boldsymbol{r}).
\end{flalign}
As shown in FIG.~\ref{fig:_n(r)}, $\mathbbm{n}(\boldsymbol{r})$ has the geometric interpretation as measuring the overlap area between region $\textrm{B}$ and region $\textrm{D}$ after translating region $\textrm{D}$ by $\boldsymbol{r}$. It is not hard to show that the area of the region is
\begin{flalign}
    \mathbbm{n}(\boldsymbol{r})=r^{2}\frac{\sin(\phi-\varphi)\sin(\theta-\varphi+\phi)}{\sin(\theta)}.
\end{flalign}
The generalized universal angle functions in Eq.~\eqref{eq:_angle_functions} are obtained from the following integrals
\begin{flalign}
f_{n}(\theta)=2\int_{\phi+\theta}^{\phi+\pi}\textrm{d}\varphi\frac{\sin(\phi-\varphi)\sin(\theta-\varphi+\phi)}{\sin(\theta)}e^{\mathtt{i}nN(\varphi-\phi)}.
\end{flalign}

\begin{figure}[t]
\centering
\begin{tikzpicture}[scale=0.6]
\draw[dashed, ->] (-5,0) -- (5,0) node[anchor=north west] {$x$};
\draw[dashed, ->] (0,-5) -- (0,5) node[anchor=south east] {$y$};

\draw[thick] (-5,-1.5) -- (5,1.5);
\draw[thick] (-2,-5) -- (2,5);
\node at (-3,3) {$B$};
\node at (3,-3) {$D$};

\draw[thick, dashed] (-5,0.1) -- (3,2.5);
\draw[thick, dashed] (-4,-4) -- (-0.5,4.75);
\node at (-1.1,0.3) {$\boldsymbol{r}$};

\fill[blue, opacity=0.3] (0,0) -- (0.727273,1.818182) -- (-2,1) -- (-2.727273,-0.818182) -- cycle;

\draw[thick, ->, >=stealth, line width=1.2pt] (0,0) -- (-2,1);
\draw (0.5,0) arc[start angle=0,end angle=16.7,radius=0.5];
\node at (2,0.3) {$\phi$};
\draw (0.3,0.09) arc[start angle=16.7,end angle=68.2,radius=0.3];
\node at (0.5,0.5) {$\theta$};
\end{tikzpicture}
\caption{Geometric interpretation of $\mathbbm{n}(\boldsymbol{r})$ as the area of the blue region.}
\label{fig:_n(r)}
\end{figure}
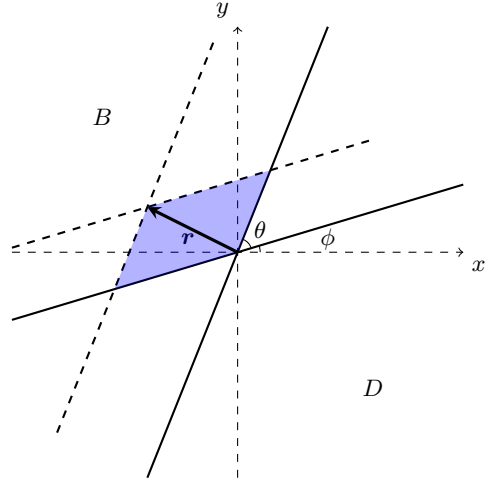

\subsection{Role of Quantum Metric}

We are interested in the role of the quantum metric $\mathsf{S}_{2}^{ab}$ in the expansion Eq.~\eqref{eq:_BF_corner_expd} of corner charge fluctuations. We first examine its contribution to different $S_{n}(q)$ in Eq.~\eqref{eq:_SSF_3}. By substituting $\mathsf{S}_{2}^{ab}q_{a}q_{b}$ into Eq.~\eqref{eq:_SSF_3_coef}, we find that
\begin{flalign}
\int_{0}^{2\pi}\frac{\textrm{d}\varphi}{2\pi}(\mathsf{S}_{2}^{ab}q_{a}q_{b})e^{-\mathtt{i}m\varphi}=\left(\delta_{0,m}\frac{\mathsf{S}_{2}^{xx}+\mathsf{S}_{2}^{yy}}{2}+\delta_{+2,m}\frac{\mathsf{S}_{2}^{xx}-\mathsf{S}_{2}^{yy}-\mathtt{i}2\mathsf{S}_{2}^{xy}}{4}+\delta_{-2,m}\frac{\mathsf{S}_{2}^{xx}-\mathsf{S}_{2}^{yy}+\mathtt{i}2\mathsf{S}_{2}^{xy}}{4}\right)q^{2},
\end{flalign}
where $(q_{x},q_{y})=q(\cos\varphi,\sin\varphi)$ and $m=nN$. There are two interesting scenarios to consider: {\it (1)} $n=0$ for arbitrary $N\geq1$, and {\it (2)} $n=\pm1$ for $N=2$. We then obtain the values of $\mathsf{C}_{n}$ using the following trick. Utilizing the properties of Bessel functions, one can verify that  
\begin{flalign}
\lim_{q\rightarrow0}\frac{\textrm{d}^{2}}{\textrm{d}q^{2}}J_{m}(qr)=\lim_{q\rightarrow0}\frac{r^{2}}{4}(J_{m-2}(qr)-2J_{m}(qr)+J_{m+2}(qr))=-\frac{r^{2}}{2}\delta_{0,m}+\frac{r^{2}}{4}\delta_{+2,m}+\frac{r^{2}}{4}\delta_{-2,m},
\end{flalign}
where $m=nN$. Based on the Fourier relations in Eq.~\eqref{eq:_radial_Fourier}, the two interesting scenarios satisfy
\begin{flalign}
N\geq1:\quad&\lim_{q\rightarrow0}\frac{\textrm{d}^{2}}{\textrm{d}q^{2}}S_{0}(q)=-2\pi\int_{0}^{+\infty}\textrm{d}r\frac{r^{3}}{2}S_{0}(r),\nonumber\\N=2:\quad&\lim_{q\rightarrow0}\frac{\textrm{d}^{2}}{\textrm{d}q^{2}}S_{\pm1}(q)=-2\pi\int_{0}^{+\infty}\textrm{d}r\frac{r^{3}}{4}S_{\pm1}(r).
\label{eq:_2nd_derv_Sq}
\end{flalign}

\subsection{High-Order Harmonics}

For illustration purposes, we will discuss the case where $N=2$. The results for $N \geq 3$ can be easily derived from the $N=2$ case by relabeling any coefficient labeled by $n$ as $2n/N$.

We assume the static structure factor of an insulator generally has the following expansion
\begin{flalign}
S(\boldsymbol{q})=\mathsf{S}_{2}^{ab}q_{a}q_{b}+\mathsf{S}_{4}^{abcd}q_{a}q_{b}q_{c}q_{d}+\mathsf{S}_{6}^{abcdef}q_{a}q_{b}q_{c}q_{d}q_{e}q_{f}+\ldots,
\end{flalign}
where $\mathsf{S}_{2}^{ab}$ represents the quantum metric, $\mathsf{S}_{4}^{abcd},\mathsf{S}_{6}^{abcdef},\ldots$ are higher-order symmetric tensors. After plugging $S(\boldsymbol{q})$ into Eq.~\eqref{eq:_SSF_3_coef}, we find a series of radial functions 
\begin{flalign}
S_{n}(q)=u_{n1}q^{2}+u_{n2}q^{4}+u_{n3}q^{6}+u_{n4}q^{8}+\ldots
\end{flalign}
As shown previously, we have $u_{0,1}=(\mathsf{S}_{2}^{xx}+\mathsf{S}_{2}^{yy})/2$ and $u_{\pm1,1}=(\mathsf{S}_{2}^{xx}-\mathsf{S}_{2}^{yy}\mp\mathtt{i}2\mathsf{S}_{2}^{xy})/4$. All the other coefficients $u_{nm}$ can be obtained similarly. Notably, $u_{nm}\neq0$ only when $m\geq |n|$. The precise values of these coefficients are not crucial for our subsequent discussion. 

For regularization purposes, we rewrite $S_{n}(q)$ as
\begin{flalign}
S_{n}(q)&=u_{n1}(1-e^{-q^{2}})+\left(u_{n2}+\frac{u_{n1}}{2}\right)(1-e^{-q^{4}})+\left(u_{n3}-\frac{u_{n1}}{6}\right)(1-e^{-q^{6}})\nonumber\\&\ \ \ \ +\left(u_{n4}+\frac{7u_{n1}}{24}+\frac{u_{n2}}{2}\right)(1-e^{-q^{8}})+\ldots
\end{flalign}
We aim to understand the coefficient Eq.~\eqref{eq:_angle_functions_coef} through the following expansion
\begin{flalign}
\mathsf{C}_{n}=u_{n1}\mathsf{C}_{n1}+\left(u_{n2}+\frac{u_{n1}}{2}\right)\mathsf{C}_{n2}+\left(u_{n3}-\frac{u_{n1}}{6}\right)\mathsf{C}_{n3}+\left(u_{n4}+\frac{7u_{n1}}{24}+\frac{u_{n2}}{2}\right)\mathsf{C}_{n4}+\ldots
\end{flalign}
We adopt a regularization scheme similar to dimensional regularization and derive the general expression for $\mathsf{C}_{nm}$
\begin{flalign}
\mathsf{C}_{nm}&=\int_{0}^{+\infty}\textrm{d}r\frac{-r^{\alpha}}{2}\int_{0}^{+\infty}\frac{\textrm{d}q}{2\pi(-1)^{n}}q(-e^{-q^{2m}})J_{2n}(qr)\nonumber\\&=\int_{0}^{+\infty}\frac{\textrm{d}q}{2\pi(-1)^{n}}q(e^{-q^{2m}})\frac{2^{\alpha-1}\Gamma(\frac{2n+1+\alpha}{2})}{q^{\alpha+1}\Gamma(\frac{2n+1-\alpha}{2})}=\frac{2^{\alpha-3}\Gamma(\frac{1-\alpha}{2m})\Gamma(\frac{2n+1+\alpha}{2})}{(-1)^{n}\pi m\Gamma(\frac{2n+1-\alpha}{2})},
\end{flalign}
where the limit $\alpha\rightarrow3$ will be taken at the end of the calculation. The results can be summarized as follows
\begin{flalign}
\mathsf{C}_{01}&=\frac{1}{\pi},\qquad\mathsf{C}_{02}=\mathsf{C}_{03}=\mathsf{C}_{04}=\ldots=0,\nonumber\\\mathsf{C}_{11}&=\frac{2}{\pi},\qquad\mathsf{C}_{12}=\mathsf{C}_{13}=\mathsf{C}_{14}=\ldots=0,\nonumber\\\mathsf{C}_{nm}&=\frac{(-1)^{n+1}n(n^{2}-1)\Gamma(\frac{m-1}{m})}{\pi}\neq0\textrm{ for all }m\geq n\textrm{ where }n\geq2.
\end{flalign}
Additionally, when $n<0$, one has $\mathsf{C}_{n,m}=\mathsf{C}_{-n,m}$.

For $n=0$ and $n=\pm1$, we find that the coefficients in Eq.~\eqref{eq:_angle_functions_coef} are fully determined by the quantum metric 
\begin{flalign}
\mathsf{C}_{0}&=\frac{u_{0,1}}{\pi}=\frac{\mathsf{S}_{2}^{xx}+\mathsf{S}_{2}^{yy}}{2\pi},\nonumber\\\mathsf{C}_{\pm1}&=\frac{2u_{\pm1,1}}{\pi}=\frac{\mathsf{S}_{2}^{xx}-\mathsf{S}_{2}^{yy}\mp\mathtt{i}2\mathsf{S}_{2}^{xy}}{2\pi},
\end{flalign}
which are consistent with Eq.~\eqref{eq:_2nd_derv_Sq}. When $n\geq2$, each $\mathsf{C}_{n}$ is represented as the summation of an infinite series.

\section{Generalized Kohn Theorem} \label{app:_Kohn_th}

In this Appendix, we present two complementary approaches to prove the generalized Kohn theorem for systems that are invariant under Galilean boosts and continuous translation, without assuming continuous rotational symmetry. The first one is based on the microscopic Hamiltonian in Eq.~\eqref{eq:_Kohn_Hamiltonian}, while the second relies solely on Ward identities.

\subsection{Hamiltonian Approach}

We extend the original Kohn theorem~\cite{Kohn1964} (also see Ref.~\cite{Zhang1992CSLG}) to anisotropic
Landau levels described by the Hamiltonian Eq.~\eqref{eq:_Kohn_Hamiltonian}. The canonical commutation relations between $\boldsymbol{p}^{j}$ and $\boldsymbol{r}_{j}$ lead to  
\begin{equation}
[r_{i}^{a},\pi_{b}^{j}]=\mathtt{i}\delta_{i}^{j}\delta_{b}^{a},\qquad[\pi_{a}^{i},\pi_{b}^{j}]=\mathtt{i}\delta^{ij}\varepsilon_{ab} B,
\end{equation}
where $\varepsilon_{ab}=\varepsilon^{ab}$ denotes the Levi-Civita
symbol, and $\ell_{B}=\sqrt{1/|B|}$ is the magnetic length. Following Ref.~\cite{Qiu2012AniQH}, we decompose $\eta_{ab}$ and $\eta^{ab}$ in terms
of complex vectors $\zeta_{a}$ and $\lambda^{a}$
\begin{equation}
\eta_{ab}=\zeta^{*}_{a}\zeta_{b}+\zeta^{*}_{b}\zeta_{a},\qquad\eta^{ab}=\lambda^{*a}\lambda^{b}+\lambda^{*b}\lambda^{a}.
\end{equation}
In terms of the components of the unimodular metric, we can express the complex vector $\zeta_{a}$ as
\begin{equation}
(\zeta_{x},\zeta_{y})=\frac{1}{\sqrt{2}}(\sqrt{\eta_{xx}},\sqrt{\eta_{yy}}e^{\mathtt{i}\gamma \chi}),
\end{equation}
where $\gamma=\arccos(\eta_{xy}/\sqrt{\eta_{xx}\eta_{yy}})$ and $\chi = \mathrm{sgn}(B)$. The complex vector  $\lambda^{a}$
is related to $\zeta_{a}$ via
\begin{equation}
\lambda^{a}=-\mathtt{i} \chi\varepsilon^{ab}\zeta_{b},\qquad\lambda^{*a}=+\mathtt{i}\chi\varepsilon^{ab}\zeta^{*}_{b},\qquad\zeta_{a}=-\mathtt{i}\chi\varepsilon_{ab}\lambda^{b},\qquad\zeta^{*}_{a}=+\mathtt{i}\chi\varepsilon_{ab}\lambda^{*b}.
\end{equation}
It is straightforward to verify the following properties
\begin{flalign}
 & \im(\zeta^{*}_{a}\zeta_{b})=\frac{\chi}{2}\varepsilon_{ab},\qquad\im(\lambda^{*a}\lambda^{b})=\frac{\chi}{2}\varepsilon^{ab},\nonumber \\
 & \re (\zeta^{*}_{a}\lambda^{b})=\frac{1}{2}\delta_{a}^{b},\qquad\re(\lambda^{*a}\zeta_{b})=\frac{1}{2}\delta_{b}^{a},\nonumber \\
 & \zeta^{*}_{a}\lambda^{a}=\zeta_{a}\lambda^{*a}=1,\qquad\zeta_{a}\lambda^{a}=\zeta^{*}_{a}\lambda^{*a}=0.
\end{flalign}
In the isotropic case $\eta_{ab}=\delta_{ab}$, one has $\zeta=\lambda=\frac{1}{\sqrt{2}}(1,\mathtt{i} \chi)$. We introduce the Landau orbit ladder operators as
\begin{equation}
\bar\alpha=\frac{\ell_{B}\lambda^{a}}{\sqrt{N_{\textrm{e}}}}\sum_{j=1}^{N_{\textrm{e}}}\pi_{a}^{j},\qquad\bar\alpha^{\dagger}=\frac{\ell_{B}\lambda^{*a}}{\sqrt{N_{\textrm{e}}}}\sum_{j=1}^{N_{\textrm{e}}}\pi_{a}^{j},\label{eq:ladder_operators}
\end{equation}
where $N_e$ is the particle number. These operators satisfy the commutation relations
\begin{flalign}
[\bar\alpha,\bar\alpha^{\dagger}] & =\frac{\ell_{B}^{2}}{N_{\textrm{e}}}\sum_{i,j=1}^{N_{\textrm{e}}}\lambda^{a}\lambda^{*b}[\pi_{a}^{i},\pi_{b}^{j}]=\frac{\chi}{N_{\textrm{e}}}\sum_{i,j=1}^{N_{\textrm{e}}}\lambda^{a}\lambda^{*b}\mathtt{i}\delta^{ij}\varepsilon_{ab}=  \zeta^{*}_{a}\lambda^{a}=1,\nonumber\\{}
[H,\bar\alpha] & =\frac{\lambda^{*a}\lambda^{b}+\lambda^{*b}\lambda^{a}}{2m}\frac{\ell_{B}\lambda^{c}}{\sqrt{N_{\textrm{e}}}}\sum_{i,j=1}^{N_{\textrm{e}}}[\pi_{a}^{i}\pi_{b}^{i},\pi_{c}^{j}] = \frac{\lambda^{*a}\lambda^{b}+\lambda^{*b}\lambda^{a}}{m\ell_{B}}\frac{(-\zeta_{b})}{\sqrt{N_{\textrm{e}}}}\sum_{j=1}^{N_{\textrm{e}}}\pi_{a}^{j}=-\omega_{c}\bar\alpha,\nonumber\\{}
[H,\bar\alpha^{\dagger}] & =\frac{\lambda^{*a}\lambda^{b}+\lambda^{*b}\lambda^{a}}{2m}\frac{\ell_{B}{\lambda}^{*c}}{\sqrt{N_{\textrm{e}}}}\sum_{i,j=1}^{N_{\textrm{e}}}[\pi_{a}^{i}\pi_{b}^{i},\pi_{c}^{j}] = \frac{\lambda^{*a}\lambda^{b}+\lambda^{*b}\lambda^{a}}{m\ell_{B}}\frac{(+\zeta^{*}_{b})}{\sqrt{N_{\textrm{e}}}}\sum_{j=1}^{N_{\textrm{e}}}\pi_{a}^{j}=+\omega_{c}\bar\alpha^{\dagger},
\end{flalign}
where $\omega_{c}=1/(m\ell_{B}^{2})$ is the cyclotron frequency.
The interaction term in Eq.~\eqref{eq:_Kohn_Hamiltonian} conserves
the total momentum and therefore commutes with both $\bar\alpha$ and
$\bar\alpha^{\dagger}$. It is evident that the center-of-mass degree of freedom forms its own Landau levels and is insensitive to interactions.

\subsection{Density Response}

In the following, we examine the electromagnetic responses $\Pi^{\mu\nu}(q)$
of the system. According to the Ward identity for charge conservation, the (retarded) density response $\Pi^{tt}$ can be related to the current response
$\Pi^{ab}$ as follows
\begin{equation}
\Pi^{tt}(\omega,\boldsymbol{q})=\frac{q_{a}q_{b}}{\omega^{2}}\Pi^{ab}(\omega,\boldsymbol{q}),
\label{eq:_Ward_id_PI}
\end{equation}
From the Hamiltonian in Eq.~\eqref{eq:_Kohn_Hamiltonian}, the response current operator is given by
\begin{equation}
J^{a}(\boldsymbol{r}_{j})=-\frac{\partial H}{\partial A_{a}(\boldsymbol{r}_{j})}=\frac{\lambda^{*a}\lambda^{b}+\lambda^{*b}\lambda^{a}}{m}\pi_{b}^{j}.
\end{equation}
The Fourier transform of this operator, in the small-$\boldsymbol{q}$ expansion, is
\begin{equation}
J^{a}(\boldsymbol{q})=\sum_{j=1}^{N_{\textrm{e}}}J^{a}(\boldsymbol{r}_{j})e^{\mathtt{i}\boldsymbol{q}\cdot\boldsymbol{r}_{j}}=\sqrt{N_{\textrm{e}}}\ell_{B}\omega_{c}(\lambda^{*a}\alpha+\lambda^{a}\alpha^{\dagger})+\ldots
\end{equation}
Thus, the density response in the long-wavelength limit is
\begin{flalign}
\Pi^{tt}(\omega,\boldsymbol{q}) & =q_{a}q_{b}\frac{1}{V}\sum_{n,m}\frac{\langle\psi_{n}|J_{\boldsymbol{q}}^{a}|\psi_{m}\rangle\langle\psi_{m}|J_{-\boldsymbol{q}}^{b}|\psi_{n}\rangle}{\mathcal{Z}(E_{n}-E_{m})^{2}}\frac{e^{-\beta E_{n}}-e^{-\beta E_{m}}}{\omega+E_{n}-E_{m}+\mathtt{i}0^{+}}\nonumber \\
 & =\frac{|\nu|}{2\pi}\sum_{n,m}\frac{e^{-\beta E_{n}}-e^{-\beta E_{m}}}{\mathcal{Z}}\frac{\omega_{c}^{2}}{(E_{n}-E_{m})^{2}}\frac{|\langle\psi_{n}|q_{a}(\lambda^{*a}\alpha+\lambda^{a}\alpha^{\dagger})|\psi_{m}\rangle|^{2}}{\omega+E_{n}-E_{m}+\mathtt{i}0^{+}}+\ldots
\end{flalign}
Here, $|\psi_{n}\rangle$ are the complete eigenstates of the Hamiltonian in Eq.~\eqref{eq:_Kohn_Hamiltonian}, with $H|\psi_{n}\rangle=E_{n}|\psi_{n}\rangle$.
The partition function is $\mathcal{Z}=\sum_{n}e^{-\beta E_{n}}$, and the filling factor $\nu$ satisfies $|\nu|=2\pi\ell_{B}^{2}N_{\textrm{e}}/V$, where $V$ is the volume of the system. Due to the properties of the
ladder operators, only the terms with $E_{n}-E_{m}=\pm\omega_{c}$
contribute. At zero temperature ($\beta\rightarrow+\infty$), the long-wavelength result has the following real and imaginary parts 
\begin{flalign}
\re\Pi^{tt}(\omega,\boldsymbol{q}) & =q_{a}q_{b}\frac{|\nu|}{2\pi}\frac{\lambda^{*a}\lambda^{b}+\lambda^{*b}\lambda^{a}}{2}\frac{2\omega_{c}}{\omega^{2}-\omega_{c}^{2}}+\ldots\nonumber \\
\im\Pi^{tt}(\omega,\boldsymbol{q}) & =q_{a}q_{b}\frac{|\nu|}{2\pi}\frac{\lambda^{*a}\lambda^{b}+\lambda^{*b}\lambda^{a}}{2}(\pi\delta(\omega+\omega_{c})-\pi\delta(\omega-\omega_{c}))+\ldots
\end{flalign}
This is valid for systems with degenerate or non-degenerate ground states. For the isotropic case, we have $\lambda^{*a}\lambda^{b}+\lambda^{*b}\lambda^{a}=\delta^{ab}$, and the standard Kohn theorem~\cite{Kohn1964} is recovered. 

The quadratic term in the static structure factor $S(\boldsymbol{q})=\mathsf{S}_{2}^{ab}q_{a}q_{b}+\ldots$
is therefore given by
\begin{equation}
\mathsf{S}_{2}^{ab}=\frac{|\nu|}{4\pi}(\lambda^{*a}\lambda^{b}+\lambda^{*b}\lambda^{a})=\frac{|\nu|}{4\pi}\eta^{ab},
\end{equation}
which is determined by the filling factor $|\nu|$ and the unimodular metric $\eta^{ab}$. {The fact that the many-body quantum metric $\mathsf{S}_{2}^{ab}$ is not renormalized by interactions can also be understood from the decoupling of the center-of-mass coordinate $\bar{\boldsymbol{r}}=(1/N_{\textrm{e}})\sum_{j}\boldsymbol{r}_{j}$, since the static structure factor can be expanded as $S(\boldsymbol{q})\sim\langle\rho(\boldsymbol{q})\rho(-\boldsymbol{q})\rangle-\langle\rho(\boldsymbol{q})\rangle\langle\rho(-\boldsymbol{q})\rangle\sim q_{a}q_{b}(\langle\bar{r}^{a}\bar{r}^{b}\rangle-\langle\bar{r}^{a}\rangle\langle\bar{r}^{b}\rangle)+\ldots$ where $\rho(\boldsymbol{q})=\sum_{j}e^{\mathtt{i}\boldsymbol{q}\cdot\boldsymbol{r}_{j}}$ is the density operator.}


\subsection{Ward-Identity Approach}

The generalized Kohn theorem can also be understood through symmetries and Ward identities. Let us extend the discussion in Ref.~\cite{hoyos2014rev} to systems in flat spacetime with an anisotropic mass tensor $m\eta_{ab}$. Invariance under Galilean boosts connects the momentum density $T_{ta}$ to the current density through the relation
\begin{equation}
T_{ta}=m\eta_{ab}J^{b},
\label{eq:_Ward_id_boost}
\end{equation}
where $T_{\mu\nu}$ represents the stress tensor. In the presence of an external electromagnetic field $F_{\mu\nu}$, the Ward identity for momentum conservation is given by 
\begin{equation}
\partial^{\mu}T_{\mu a}=F_{a\nu}J^{\nu}.
\label{eq:_Ward_id_trans}
\end{equation}
The combination of Eq.~\eqref{eq:_Ward_id_boost} and Eq.~\eqref{eq:_Ward_id_trans} leads to the Ward identity
\begin{equation}
\partial^{b}T_{ba}=F_{at}J^{t}+F_{ab}J^{b}-m\eta_{ab}\partial^{t}J^{b},
\label{eq:_Ward_id_Galilean}
\end{equation}
where $F_{ta}=-E_{a}$ and $F_{ab}=B\varepsilon_{ab}$. After performing the Fourier transform to frequency-momentum space and setting the momentum to zero, we have
\begin{flalign}
&0=-\frac{1}{m}E_{a}(\omega)J^{t}(\boldsymbol{x})+(\omega_{c}\varepsilon_{ab}+\mathtt{i}\omega\eta_{ab})J^{b}(\omega)\nonumber\\\Longrightarrow\quad&J^{a}(\omega)=\frac{J^{t}(\boldsymbol{x})}{m}\frac{-\mathtt{i}\omega\eta^{ab}+\omega_{c}\varepsilon^{ab}}{\omega^{2}-\omega_{c}^{2}}E_{b}(\omega),
\end{flalign}
where we have used $\eta^{ab}\eta_{bc}=\delta_{c}^{a}$ and $\det(\eta_{ab})=1$. On the one hand, the optical conductivity can be derived from the response of $\langle J^{a}(\omega)\rangle$ to $E_{b}(\omega)$. On the other hand, it is related to the density response through the U(1) Ward identity in  Eq.~\eqref{eq:_Ward_id_PI}. Consequently, the density response satisfies
\begin{flalign}
\re\Pi^{tt}(\omega,\boldsymbol{q})=q_{a}q_{b}\frac{\langle J^{t}\rangle}{m}\frac{\eta^{ab}}{\omega^{2}-\omega_{c}^{2}}+\mathscr{O}(|\boldsymbol{q}|^{3})=q_{a}q_{b}\eta^{ab}\frac{|\nu|}{2\pi}\frac{\omega_{c}}{\omega^{2}-\omega_{c}^{2}}+\mathscr{O}(|\boldsymbol{q}|^{3}),
\end{flalign}
where $\nu=(2\pi/B)\langle J^{t}\rangle$ is the filling factor. Furthermore, the value of $\im\Pi^{tt}(\omega,\boldsymbol{q})$ can be determined from $\textrm{Re}\Pi^{tt}(\omega,\boldsymbol{q})$ using the Kramers-Kronig relations. Note that the Ward identity in Eq.~\eqref{eq:_Ward_id_Galilean} contains much more information than just the Kohn theorem. Specifically, it can also be used to relate Hall conductivity and Hall viscosity in anisotropic systems by extending the discussion in Ref.~\cite{hoyos2014rev}.

\section{Generalized Rotational Invariance}
\label{app:_QM_Sym}

In this appendix, we examine how bipartite fluctuations behave under coordinate transformations. To place our discussion in a broader context, we assume only the generalized rotational invariance, where the static structure factor $S(\boldsymbol{q})$ is a function of the ``quantum distance'' $g^{ab}q_{a}q_{b}$. This approach does not require invariance under Galilean boosts, making it more general than Landau levels described by Eq.~\eqref{eq:_Kohn_Hamiltonian}.

We begin by recalling that bipartite fluctuations can be calculated as
\begin{flalign}
\mathcal{N}_{\textrm{A}}^{[2]}=\int\frac{\textrm{d}^{2}\boldsymbol{q}}{(2\pi)^{2}}S(\boldsymbol{q})|\Theta_{\textrm{A}}(\boldsymbol{q})|^{2},
\label{eq:_BF_2}
\end{flalign}
where the spatial geometry dependence is encapsulated in the function
\begin{flalign}
\Theta_{\textrm{A}}(\boldsymbol{q})=\int_{\textrm{A}}\textrm{d}^{2}\boldsymbol{x}e^{\mathtt{i}\boldsymbol{q}\cdot\boldsymbol{x}}.
\end{flalign}
For a gapped insulator, the Taylor expansion of the static structure factor can generally be expressed as 
\begin{flalign}
S(\boldsymbol{q})=g^{ab}q_{a}q_{b}+\sum_{m=2}^{+\infty}a_{m}(g^{ab}q_{a}q_{b})^{m},
\end{flalign}
where $a_{m}$ are a series of constants. We denote the inverse of the quantum metric by $g_{ab}$ satisfying $g_{ab}g^{bc}=\delta_{a}^{c}$. Then we introduce a set of vielbeins, or frame fields, $e_{a}^{I}$ to diagonalize the quantum metric 
\begin{flalign}
g_{ab}=e_{a}^{I}e_{b}^{J}\delta_{IJ}.
\end{flalign}
It would be more convenient to work in a new coordinate system defined by
\begin{flalign}
\tilde{x}^{I}=e_{a}^{I}x^{a},\qquad\tilde{q}_{I}=e_{I}^{a}q_{a},
\end{flalign}
where $e_{I}^{a}$ is defined such that $e_{a}^{I}e_{I}^{b}=\delta_{a}^{b}$. Now, Eq.~\eqref{eq:_BF_2} can be carefully rewritten as  
\begin{flalign}
\mathcal{N}_{\textrm{A}}^{[2]}&=\frac{1}{\sqrt{\textrm{det}(g^{ab})}}\int\frac{\textrm{d}^{2}\tilde{\boldsymbol{q}}}{(2\pi)^{2}}\tilde{S}(\tilde{\boldsymbol{q}})|\tilde{\Theta}_{\textrm{A}}(\tilde{\boldsymbol{q}})|^{2},\nonumber\\\tilde{\Theta}_{\textrm{A}}(\tilde{\boldsymbol{q}})&=\sqrt{\textrm{det}(g^{ab})}\int_{\tilde{\textrm{A}}}\textrm{d}^{2}\tilde{\boldsymbol{x}}e^{\mathtt{i}\tilde{\boldsymbol{q}}\cdot\tilde{\boldsymbol{x}}},\nonumber\\\tilde{S}(\tilde{\boldsymbol{q}})&=\delta^{IJ}\tilde{q}_{I}\tilde{q}_{J}+\sum_{m=2}^{+\infty}a_{m}(\delta^{IJ}\tilde{q}_{I}\tilde{q}_{J})^{m}
\label{eq:_BF_3}
\end{flalign}
where $\tilde{\textrm{A}}$ represents the real-space subregion after the coordinate transformation. The Jacobians of the integrals satisfy $\det(e_{I}^{a})=1/\det(e_{a}^{I})=\sqrt{\textrm{det}(g^{ab})}$. Compared to Eq.~\eqref{eq:_BF_2}, the final result in Eq.~\eqref{eq:_BF_3} has the form 
\begin{flalign}
\frac{\mathcal{N}_{\textrm{A}}^{[2]}}{\sqrt{\textrm{det}(g^{ab})}}=\int\frac{\textrm{d}^{2}\tilde{\boldsymbol{q}}}{(2\pi)^{2}}\tilde{S}(\tilde{\boldsymbol{q}})\left|\int_{\tilde{\textrm{A}}}\textrm{d}^{2}\tilde{\boldsymbol{x}}e^{\mathtt{i}\tilde{\boldsymbol{q}}\cdot\tilde{\boldsymbol{x}}}\right|^{2}.
\end{flalign}
The right-hand side describes the bipartite fluctuations associated with a subregion $\tilde{\textrm{A}}$, for a system with the static structure factor $\tilde{S}(\tilde{\boldsymbol{q}})$. The gapped modes of this new system are isotropic, and according to calculations in Ref.~\cite{uni_corner}, they contribute to a corner term of $(-1/\pi)(1+(\pi-\tilde{\theta})\cot\tilde{\theta})$, where $\tilde{\theta}$ is the corner angle of $\tilde{\textrm{A}}$. 

In conclusion, the bipartite fluctuations of the original system are given by
\begin{flalign}
\mathcal{N}_{\textrm{A}}^{[2]}(\phi,\theta)=\#L-\frac{\sqrt{\det(g^{ab})}}{\pi}f_{0}(\tilde{\theta}(\phi,\theta))
\end{flalign}
where $f_{0}$ is given by Eq.~\eqref{eq:_angle_function_0} (i.e., the zeroth angle function $f_{0}^{c}$ in Eq.~\eqref{eq:_angle_functions}). The coefficient $\sqrt{\det(g^{ab})}$ has the geometric interpretation as quantum volume, which satisfies the determinant bounds of Eq.~\eqref{eq:_detS2_bounds}. The new angle variable $\tilde{\theta}$ can be determined from the original angles $\theta$ and $\phi$ using the relation
\begin{flalign}
\cos\tilde{\theta}=\frac{g_{ab}\hat{n}^{a}\hat{m}^{b}}{\sqrt{g_{ab}\hat{n}^{a}\hat{n}^{b}}\sqrt{g_{ab}\hat{m}^{a}\hat{m}^{b}}},
\end{flalign}
where $\hat{n}=(\cos\phi,\sin\phi)$ and $\hat{m}=(\cos(\phi+\theta),\sin(\phi+\theta))$. Concerning the small-$\theta$ singularity of the corner term discussed in the main text, it can be shown that  
\begin{flalign}
f_{0}(\tilde{\theta}(\phi,\theta\rightarrow0))\approx\frac{\pi}{\theta}\frac{\sin^{2}(\phi)g^{xx}+\cos^{2}(\phi)g^{yy}-\sin(2\phi)g^{xy}}{\sqrt{\det(g^{ab})}}.
\end{flalign}
Additionally, for the orientation-averaged corner term, we have numerically verified that
\begin{flalign}
\int_{0}^{2\pi}\frac{\textrm{d}\phi}{2\pi}\frac{\sqrt{\det(g^{ab})}}{\pi}f_{0}(\tilde{\theta}(\phi,\theta))=\frac{\textrm{Tr}(g^{ab})}{2\pi}f_{0}(\theta).
\end{flalign}

\section{Response Theory and Quantum Geometry} 
\label{app:_Quan_Geom}

In this Appendix, we briefly review the standard relations~\cite{SWM2000,Resta2011Rev} between linear response theory and many-body quantum geometry. To formally express the Källén-Lehmann spectral representation of response functions, we introduce the complete eigenstates $|\psi_{n}\rangle$ of a many-body Hamiltonian $H$ in $d$ spatial dimensions, such that $H|\psi_{n}\rangle=E_{n}|\psi_{n}\rangle$. To simplify the notation, we also define 
\begin{flalign}
\mathcal{R}_{nm}^{ab}&=\frac{1}{V}\re[\langle\psi_{n}|\mathcal{J}^{a}|\psi_{m}\rangle\langle\psi_{m}|\mathcal{J}^{b}|\psi_{n}\rangle],\nonumber\\\mathcal{I}_{nm}^{ab}&=\frac{1}{V}\im[\langle\psi_{n}|\mathcal{J}^{a}|\psi_{m}\rangle\langle\psi_{m}|\mathcal{J}^{b}|\psi_{n}\rangle],
\end{flalign}
where $\mathcal{J}^{a}=\int\textrm{d}^{d}\boldsymbol{x}J^{a}(t=0,\boldsymbol{x})=\int\textrm{d}^{d}\boldsymbol{x}(-\partial H/\partial A_{a})$ is the integrated current operator, and $V=\int\textrm{d}^{d}\boldsymbol{x}$ represents the total spatial volume of the system. 

%
%

The longitudinal conductivity $\sigma_{+}^{ab}=(\sigma^{ab}+\sigma^{ba})/2$ can be written as 
\begin{flalign}
\sigma_{+}^{ab}(\omega)=D^{ab}\left(\delta(\omega)+\frac{\mathtt{i}}{\pi\omega}\right)+\sigma_{\textrm{reg}}^{ab}(\omega).
\label{eq:_conductivity_long}
\end{flalign}
The Drude weight, which is transpose-symmetric ($D^{ab}=D^{ba}$), is obtained as 
\begin{flalign}
D^{ab}=\lim_{A\rightarrow0}\frac{\pi}{V}\left\langle \!\frac{\partial^{2}H}{\partial A_{a}\partial A_{b}}\!\right\rangle +\sum_{n,m}\frac{e^{-\beta E_{n}}-e^{-\beta E_{m}}}{\mathcal{Z}}\frac{\pi\mathcal{R}_{nm}^{ab}}{E_{n}-E_{m}}.
\label{eq:_Drude}
\end{flalign}
Here, $A_{a}$ denotes the background U(1) field, $E_{n}$ are energy levels of the many-body Hamiltonian $H$, and $\mathcal{Z}=\sum_{n}e^{-\beta E_{n}}$ represents the partition function.  The regular term in Eq.~\eqref{eq:_conductivity_long} has the real and imaginary parts
\begin{flalign}
\re\sigma_{\textrm{reg}}^{ab}(\omega)&=\sum_{n,m}\frac{e^{-\beta E_{n}}-e^{-\beta E_{m}}}{\mathcal{Z}}\frac{\pi\mathcal{R}_{nm}^{ab}}{\omega}\delta(\omega+E_{n}-E_{m}),\\
\im\sigma_{\textrm{reg}}^{ab}(\omega)&=\sum_{n,m}\frac{e^{-\beta E_{n}}-e^{-\beta E_{m}}}{\mathcal{Z}}\frac{-\mathcal{R}_{nm}^{ab}}{(\omega+E_{n}-E_{m})(E_{n}-E_{m})}.
\end{flalign}
In addition, the Hall conductivity $\sigma_{-}^{ab}=(\sigma^{ab}-\sigma^{ba})/2$ has the spectral representation 
\begin{flalign}
\re\sigma_{-}^{ab}(\omega)&=\sum_{n,m}\frac{e^{-\beta E_{n}}-e^{-\beta E_{m}}}{\mathcal{Z}}\frac{\mathcal{I}_{nm}^{ab}}{\omega}\frac{-1}{\omega+E_{n}-E_{m}},\\
\im\sigma_{-}^{ab}(\omega)&=\sum_{n,m}\frac{e^{-\beta E_{n}}-e^{-\beta E_{m}}}{\mathcal{Z}}\frac{\pi\mathcal{I}_{nm}^{ab}}{\omega}\delta(\omega+E_{n}-E_{m}).
\label{eq:_conductivity_Hall}
\end{flalign}
In the literature, the absorptive part of the conductivity is defined as
\begin{flalign}
\sigma_{\textrm{abs}}^{ab}(\omega)=\frac{\sigma^{ab}(\omega)+[\sigma^{ba}(\omega)]^{*}}{2}=\textrm{Re}\sigma_{+}^{ab}(\omega)+\mathtt{i}\textrm{Im}\sigma_{-}^{ab}(\omega).
\end{flalign}

{In the following, we assume that the Drude weight vanishes, as in the cases of quantum Hall insulators and composite fermi liquids. Let us consider a $d$-dimensional periodic system with a linear size $L$ (i.e., on a $d$-dimensional torus). In the presence of background fluxes $\varPhi_{a}$ such that $A_{a}=A_a^{0}-\varPhi_{a}/L$ in the $\hat{x}_{a}$ direction. We choose a Heisenberg-like representation where change in twisted boundary condition perturbs the many-body Hamiltonian by $\delta H=-\mathcal{J}^{a} \delta A_{a}= \mathcal{J}^{a} \delta\varPhi_{a}/L$ but the eigenstates $\ket{\psi_n, \varPhi_a}$ always satisfy periodic boundary conditions. At zero temperature $\beta\rightarrow+\infty$, one can easily verify the following sum rule using the previously discussed spectral representations}
\begin{flalign}
\int_{0}^{+\infty}\frac{\textrm{d}\omega}{\pi}\frac{\sigma_{\textrm{abs}}^{ab}(\omega)}{\omega}=\frac{1}{N_{\textrm{g}}}\sum_{m=1}^{N_{\textrm{g}}}\sum_{E_n\neq E_m}\frac{\mathcal{R}_{mn}^{ab}+\mathtt{i}\mathcal{I}_{mn}^{ab}}{(E_{n}-E_{m})^{2}}=\frac{1}{L^{d-2}}\sum_{m=1}^{N_{\textrm{g}}}\frac{\mathcal{Q}_{mm}^{ab}}{N_{\textrm{g}}},
\label{eq:_QGT_sum}
\end{flalign}
where $m$ labels the possibly degenerate ground states, and $n$ runs over all energy levels. The so-called many-body quantum geometric tensor $\mathcal{Q}^{ab}$ is defined by 
\begin{flalign}
\mathcal{Q}_{mn}^{ab}=\left\langle \!\frac{\partial\psi_{m}}{\partial\varPhi_{a}}\!\right|\!(1-P)\!\left|\!\frac{\partial\psi_{n}}{\partial\varPhi_{b}}\!\right\rangle,
\end{flalign}
where $m,n$ label the ground states, and $P=\sum_{n=1}^{N_{\textrm{g}}}|\psi_{n}\rangle\langle\psi_{n}|$ denotes the projection to the ground-state manifold. It can be written as $\mathcal{Q}^{ab}=\mathcal{G}^{ab}-\frac{\mathtt{i}}{2}\mathcal{F}^{ab}$, where $\mathcal{F}^{ab}$ is the (non-abelian) Berry curvature and $\mathcal{G}^{ab}$ is the (non-abelian) quantum metric. Therefore, the localization tensor Eq.~\eqref{eq:_SWM_sum} is given by the real part of the sum rule Eq.~\eqref{eq:_QGT_sum}
\begin{flalign}
\mathsf{S}_{2}^{ab}=\int_{0}^{+\infty}\frac{\textrm{d}\omega}{\pi}\frac{\re\sigma_{\textrm{abs}}^{ab}(\omega)}{\omega}=\frac{1}{L^{d-2}}\sum_{n=1}^{N_{\textrm{g}}}\frac{\mathcal{G}_{nn}^{ab}}{N_{\textrm{g}}}.
\end{flalign}
In $d=2$, the geometric interpretation of $\mathsf{S}_{2}^{ab}$ holds independent of the system size $L$.

{It is easy to see the equivalence of this definition of localization tensor and the one in Eq. \eqref{eq:S2_from_C}. In the latter, we have used Schrodinger-like representation with $\mathcal{T}^{(L),j}_a \ket{\psi_n^S} = e^{\mathtt{i}\varPhi_a} \ket{\psi_n^S}$, where $\mathcal{T}^{(L),j}_a$ translates the $j^{\rm th}$ particle in $\hat x_a$ direction by the linear dimension of torus $L$ and we have used $S$ superscript to label Schrodinger-like representation. The transformation $\ket{\psi_n} = e^{-\mathtt{i} \varPhi_a \sum_{j=1}^{N_e}r_j^a} \ket{\psi^S_n}$ translates it to Heisenberg-like representation. This gives $L q_a = -\delta \varPhi_a$ from which Eq. \eqref{eq:S2_from_C} follows.}

\subsection{Topological Lower Bound}

For convenience, we define the geometric quantities
\begin{flalign}
Q^{ab}=\frac{1}{N_{\textrm{g}}}\sum_{n=1}^{N_{\textrm{g}}}\mathcal{Q}_{nn}^{ab},\qquad g^{ab}=\frac{1}{N_{\textrm{g}}}\sum_{n=1}^{N_{\textrm{g}}}\mathcal{G}_{nn}^{ab},\qquad\Omega^{ab}=\frac{1}{N_{\textrm{g}}}\sum_{n=1}^{N_{\textrm{g}}}\mathcal{F}_{nn}^{ab}.
\end{flalign}
When the ground state is non-degenerated (i.e., $N_{\textrm{g}}=1 $), they reduce back to the abelian quantum geometric tensor, quantum metric, and Berry curvature. Several useful inequalities can be established by utilizing the positive semi-definiteness of the complex metric $Q^{ab}=g^{ab}-\frac{\mathtt{i}}{2}\Omega^{ab}$. In $d=2$, $\textrm{det}(Q)\geq0$ is equivalent
to 
\begin{equation}
\textrm{det}(g^{ab})\geq\frac{1}{4}|\Omega^{xy}|^{2}.\label{eq:_det_bound}
\end{equation}
Furthermore, the quantum metric satisfies the inequality
\begin{equation}
(\textrm{Tr}(g^{ab}))^{2}-4\textrm{det}(g^{ab})=4g^{xy}g^{yx}+(g^{xx}-g^{yy})^{2}\geq0,\label{eq:_Tr_bound}
\end{equation}
where the bound is saturated only when $g^{ab}\propto\delta^{ab}$. Therefore, we generally have
the inequality $(\textrm{Tr}(g^{ab}))^{2}\geq4\textrm{det}(g^{ab})\geq0$.
This, combined with Eq.~\eqref{eq:_det_bound}, establishes the universal
bounds for $\mathsf{S}_{2}^{ab}$ in $d=2$
\begin{equation}
\textrm{Tr}(g^{ab})\geq2\sqrt{\textrm{det}(g^{ab})}\geq|\Omega^{xy}|\quad\Longrightarrow\quad\textrm{Tr}(\mathsf{S}_{2}^{ab})\geq2\sqrt{\textrm{det}(\mathsf{S}_{2}^{ab})}\geq|\sigma^{xy}|.
\label{eq:_low_bounds}
\end{equation}
The DC Hall conductivity $\sigma^{xy}=(2\pi)^{-2}\int_{0}^{2\pi}\textrm{d}\varPhi_{x}\int_{0}^{2\pi}\textrm{d}\varPhi_{y}\Omega^{xy}$ is obtained through the many-body Chern number.

\subsection{Energetic Upper Bound}

Based on the spectral representation of Eq.~\eqref{eq:_conductivity_long},
one can easily verify the standard $f$-sum rule 
\begin{equation}
\int_{0}^{+\infty}\frac{\textrm{d}\omega}{\pi}\re\sigma_{+}^{ab}(\omega)=\lim_{A\rightarrow0}\frac{1}{2V}\left\langle \!\frac{\partial^{2}H}{\partial A_{a}\partial A_{b}}\!\right\rangle = \frac{1}{2}\sD_2^{ab} .
\end{equation} 
For gapped phases, $\re\sigma_{+}^{ab}(\omega)$ is nonzero only above the energy gap, which we denote by $\Delta$. According to the Souza-Wilkens-Martin sum rule, we have
\begin{equation}
\mathsf{S}_{2}^{ab}=\int_{0}^{+\infty}\frac{\textrm{d}\omega}{\pi}\frac{\re\sigma_{+}^{ab}(\omega)}{\omega}\leq\int_{0}^{+\infty}\frac{\textrm{d}\omega}{\pi}\frac{\re\sigma_{+}^{ab}(\omega)}{\Delta}= \frac{1}{2\Delta}\sD_2^{ab}.
\label{eq:_up_bound}
\end{equation}
Here, the inequality means that the difference between the right-hand side and the left-hand side is a positive semi-definite matrix. This energetic upper bound applies to the localization tensor $\mathsf{S}_{2}^{ab}$ in general dimensions.

We can use this matrix inequality to derive the right side of the determinant bounds of Eq.~\eqref{eq:_detS2_bounds}. To this end, let us consider two real, symmetric and positive semi-definite matrices $M^{ab}, N^{ab}$ such that $M^{ab}-N^{ab} \geq 0$. We can rewrite $M^{ab} = \sqrt{\det(M)} \tilde M^{ab}$ where $\tilde M^{ab}$ has unit determinant. We have:
\begin{align}
   \tilde{M}^{ab}-\frac{\sqrt{\det(N)}}{\sqrt{\det(M)}}\tilde{N}^{ab}\geq0\quad\implies\quad\delta^{ab}-\frac{\sqrt{\det(N)}}{\sqrt{\det(M)}}\tilde{O}^{ab}\geq0
\end{align}
where $\tilde O=\sqrt{\tilde M^{-1}}\tilde N \sqrt{\tilde M^{-1}}$ is unimodular and positive semi-definite. Positivity of the eigenvalues on the left-hand side leads to the inequality
\begin{align}
    \frac{\sqrt{\det(M)}}{\sqrt{\det(N)}} \geq \frac{\mathrm{Tr}(\tilde O)}{2} + \sqrt{\frac{\mathrm{Tr}(\tilde O)^2}{4} - 1} \ \ \geq 1
\end{align}
where we have used $\mathrm{Tr}(\tilde O) \geq 2\sqrt{\det (\tilde O)}=2$. This then gives rise to the upper bound in Eq.~\eqref{eq:_detS2_bounds}.





\section{Polarization Fluctuations and Translational Symmetry in Landau Levels \label{app:_translational_symmetry}}
In this appendix, we discuss the translational symmetry in continuum models and its consequences on the polarization fluctuations. We focus on interacting many-body states that have non-trivial ground state degeneracy on the torus. Translations for the $j^{\rm th}$ particle are generated by the magnetic momentum
\begin{align}
    P_a^j = -B\epsilon_{ab} R^b_j
\end{align}
where $a=x,y$, $B$ is the transverse magnetic field and we have set electron charge $e=1$. Moreover, $R^a_j = r^a_j + \epsilon^{ab}\pi_b^j/B$ is the guiding center position, $r^a_j$ is the coordinate, and $\pi_a^j = p_a^j - A_a(\bm r_j)$ is the kinetic momentum for $j^{\rm th}$ particle. The magnetic momenta do not commute, i.e. $[P_a^j, P_b^k] = -i \hbar B \delta^{jk}\epsilon_{ab}$ owing to $[R^a_j, R^b_k] = -i\hbar \delta_{jk}\epsilon^{ab}/B$. Using magnetic translation operators, we can define twisted boundary conditions as follows:
\begin{align}
    e^{i P_a^j L^a/\hbar} \ket{\Psi} &= e^{i\varPhi_a} \ket{\Psi} \label{eq:twisted_bc}
\end{align}
where the indices $a, j$ are not summed over.


As discussed in main text, we generalize the generating function of moments of polarization of Ref. \cite{SWM2000} to the case where there are multiple degenerate ground states. The new generating function is an operator in the basis of ground states given by:
\begin{align}
    C_{nm}(\bm q) &= \braket{\Psi_n'|e^{-iq_a \bar r^a N_e}|\Psi_m}\nonumber\\
    \bar r^a &= \frac{1}{N_e}\sum_{j=1}^{N_e} r^a_j
\end{align}
where $N_e$ is the number of electrons. Moreover, when $q_a$ is an integer multiple of $2\pi/L^a$, $e^{-iq_a \bar r^a N_e}\ket{\Psi_n}$ obeys the same boundary conditions as $\ket{\Psi_n}$. The localization tensor can be obtained using
\begin{align}
    \mathsf{S}_{2}^{ab} &= -\left.\frac{1}{V N_{\textrm{g}}} \int \frac{\textrm{d}^2\varPhi}{(2\pi)^2}\ \frac{\partial^2 }{\partial q_a \partial q_b}\mathrm{tr}\log C(\bm q) \right|_{\bm q=0}. \label{eq:S2_Cq}
\end{align}
Using $\bar r^a = \frac{1}{N_e}\epsilon^{ab}(\bar P_b - \bar\pi_b)/B$, we can write:
\begin{gather}
    C_{nm}(\bm q) = \braket{\Psi_n'|e^{i\epsilon^{ab}q_a \bar \pi_b/B}e^{-i\epsilon^{ab}q_a \bar P_b/B}|\Psi_m}\nonumber\\
    \bar\pi_a = \sum_{j=1}^{N_e} \pi_a^j,\ \ \ \bar P_a = \sum_{j=1}^{N_e} P_a^j \label{eq:com_momenta}
\end{gather}
The first exponential operator gives rise to transitions between different Landau levels. The second operator, on the other hand, generates the translation $r^a_j \rightarrow r^a_j + \frac{B}{\hbar}\epsilon^{ab}q_b$ for each electron. We will show below that the latter does not contribute to $\mathsf{S}_{2}^{ab}$.

To deal with degenerate ground states, we notice that the minimal unitary translation operators which do not alter boundary conditions of Eq. \eqref{eq:twisted_bc} are $\mathcal{T}_y = e^{-i2\pi \bar P_y/BL^x}$ and $\mathcal{T}_x=e^{i2\pi \bar P_x/BL^y}$. 
They do not commute and satisfy $\mathcal{T}_y\mathcal{T}_x = e^{i \hbar N_e /BL_x L_y}\mathcal{T}_x\mathcal{T}_y$. 
In terms of the filling fraction $\nu \equiv hN_e/B L_x L_y$, we can write the phase factor as $e^{i2\pi \nu}$. 
Expressing $\nu=p/q$ where $p,q$ are co-prime integers with $q>0$, it is evident that $\mathcal{T}_x$ and $\mathcal{T}_y^q$ form a minimal set of operators that commute with each other and can be diagonalized simultaneously.

A translationally invariant Hamiltonian commutes with $\bar P_a$. Thus, we can find an orthonormal basis for ground state manifold labeled by $\ket{\bar k_x, \bar k_y}$ such that $\mathcal{T}_x\ket{\bar k_x, \bar k_y} = e^{i2\pi \hbar \bar k_x/BL^y} \ket{\bar k_x, \bar k_y}$, and $\mathcal{T}_y^q \ket{\bar k_x, \bar k_y} = e^{-i2\pi \hbar q \bar k_y/B L^x}\ket{\bar k_x, \bar k_y}$. Here, $\bar k_a = 2\pi \bar n_a/L^a + N_e \varPhi_a/L^a$ for $a=x,y$ and $\bar n_a \in \mathbb{Z}$ since $\mathcal{T}_a^{N_\phi} = e^{i N_e \varPhi_a}$. Since $\mathcal{T}_a$ are minimal translations, $\bar n_x \leftrightarrow \bar n_x + N_\phi$. Moreover, $\mathcal{T}_y \ket{\bar k_x, \bar k_y} = e^{-i2\pi \hbar \bar k_y/B L^x} \ket{\bar k_x-\nu BL^y/\hbar,\bar k_y}$, that amounts to $\bar n_x \rightarrow \bar n_x - N_e$ and gives rise to a state orthogonal to $\ket{\bar k_x, \bar k_y}$. Thus the action of $\mathcal{T}_y$ generates a manifold of ground states that has a dimension of ``$q$'', giving rise to a ground state degeneracy that is an integer multiple of ``$q$''.\cite{Haldane1985}

Finally, notice that a fractional power of the minimal translation modifies boundary conditions, i.e. $\ket{\Psi'} = \left(\mathcal{T}_b\right)^{\beta}\ket{\Psi}$ obeys:
\begin{align}
    e^{i P_a^j L^a/\hbar}\ket{\Psi'} = e^{i(\varPhi_a - \epsilon_{ab}2\pi\beta)} \ket{\Psi'}
\end{align}
where $a,b$ are not summed over. On the other hand, it does not alter a Hamiltonian with continuous translational invariance. Hence $\ket{\Psi'}$ is an eigenstate of the Hamiltonian if $\ket{\Psi}$ is as well, but with modified boundary conditions. In the presence of a spectral gap, $\ket{\Psi'}$ belongs to the manifold of ground states if $\ket{\Psi}$ does too. Center of mass translations are thus equivalent to flux-threading, with one flux-quantum in $a^{\rm th}$-direction corresponding to a center of mass displacement of magnitude $2\pi\hbar/BL^a$ in the direction orthogonal to $a$. Therefore, Hall conductivity $\sigma_{xy} = \nu e^2/h$, is fully determined by the filling fraction.

Armed with this machinery, we can evaluate $C_{nm}(\bm q)$ by choosing $\ket{\Psi_n'} = e^{-i\epsilon^{ab} q_a \bar P_b/B}\ket{\Psi_n}$. It is easy to see that:
\begin{align}
    \tilde C_{nm}(\bm q) &= \braket{\Psi_n|e^{i\epsilon^{ab}q_a \bar\pi_b/B}|\Psi_m}
\end{align}
where we have defined $C(\bm q) = \tilde C(\bm q)$ when the basis is chosen in the way described here. We could further choose $\ket{\Psi_n}$ to be eigenstates of $\mathcal{T}_x$ and $\mathcal{T}_y^q$, then $\tilde C_{nm}(\bm q)$ is non-zero only if $\ket{\Psi_n}$ and $\ket{\Psi_m}$ have the same eigenvalues for these operators since $[\bar\pi_a,\bar P_b] = 0$.




\end{widetext}

\bibliography{Ref}

\end{document}